\documentclass[11pt]{article}
\usepackage{epsfig} 
\newcommand{\sect}[1]{ \section{#1} \setcounter{equation}{0} }

\newcommand{\pslash}{p \! \! \! /} 
 
\newcommand{\partialslash}{\partial \! \! \! /} 
\newcommand{\half}{\mbox{\small{$\frac{1}{2}$}}} 
\newcommand{\eighth}{\mbox{\small{$\frac{1}{8}$}}} 
\newcommand{\sixteenth}{\mbox{\small{$\frac{1}{16}$}}} 
\newcommand{\threehalves}{\mbox{\small{$\frac{3}{2}$}}} 
\newcommand{\threeeighths}{\mbox{\small{$\frac{3}{8}$}}} 
\newcommand{\threesixteenths}{\mbox{\small{$\frac{3}{16}$}}} 
\newcommand{\Nc}{N_{\!c}} 
\newcommand{\Nf}{N_{\!f}} 
\newcommand{\CR}{C_2(R)} 
\newcommand{\CG}{C_2(G)} 
\newcommand{\MSbar}{\overline{\mbox{MS}}} 

\begin{document}
\title{Four loop wave function renormalization in the non-abelian Thirring 
model} 
\author{D.B. Ali \& J.A. Gracey, \\ Theoretical Physics Division, \\ Department 
of Mathematical Sciences, \\ University of Liverpool, \\ Peach Street, \\ 
Liverpool, \\ L69 7ZF, \\ United Kingdom.} 
\date{} 
\maketitle 
\vspace{5cm} 
\noindent 
{\bf Abstract.} We compute the anomalous dimension of the fermion field with 
$\Nf$ flavours in the fundamental representation of a general Lie colour group 
in the non-abelian Thirring model at four loops. The implications on the 
renormalization of the two point Green's function through the loss of 
multiplicative renormalizability of the model in dimensional regularization due
to the appearance of evanescent four fermi operators are considered at length. 
We observe the appearance of one new colour group Casimir, $d_F^{abcd} 
d_F^{abcd}$, in the final four loop result and discuss its consequences for the
relation of the Knizhnik-Zamolodchikov critical exponents in the Wess Zumino 
Witten Novikov model to the non-abelian Thirring model. Renormalization scheme 
changes are also considered to ensure that the underlying Fierz symmetry broken
by dimensional regularization is restored.  

\vspace{-20cm}
\hspace{13.5cm} 
{\bf LTH 499} 

\newpage 

\sect{Introduction.} 

The non-abelian Thirring model, (NATM), is a two dimensional renormalizable
fermionic quantum field theory with a variety of interesting properties, 
\cite{1}. For instance, it is asymptotically free, \cite{1}, and being two 
dimensional its bosonized version is related to the Wess Zumino Witten Novikov,
(WZWN), model which is a bosonic nonlinear $\sigma$ model with a topological 
term or equivalently a torsion potential, \cite{2}. Indeed its connection with 
the WZWN model has played a fundamental role in understanding two dimensional 
conformal field theories following the early study by Dashen and Frishman, 
\cite{1}, of examining the conformal properties of the NATM and its fixed point
structure. An interesting and elegant feature to emerge from the conformal 
approach to solving the NATM or equivalently the WZWN model was in the work of 
Knizhnik and Zamolodchikov, \cite{3}, where the critical exponents of the 
fields and parameters of the theory were written down exactly. These all orders 
results depended in a simple fashion on the group Casimirs of the 
underlying non-abelian symmetry of the model. In particular the elementary
Casimirs $T(R)$, $C_2(R)$ and $C_2(G)$ were involved as well as $\Nf$ the 
number of flavours if the fields had an additional internal symmetry. These
exact results of Knizhnik and Zamolodchikov have been checked in a variety of
ways. For instance, in \cite{4,5} the perturbative renormalization group 
functions were computed in the nonlinear $\sigma$ model on the group manifold
with a Wess Zumino term to several orders in perturbation theory. Then 
evaluating the expressions at the non-trivial fixed point the results were
compared with the Knizhnik-Zamolodchikov exponents when these were expanded to
the equivalent order in perturbation theory. In expressing the essence of such
a check in this condensed way it is important to recognise the technical 
difficulties in performing the renormalization of a theory with a topological
term. For example, in \cite{4,5} the use of dimensional regularization led to
the problem of loss of multiplicative renormalizability due to the appearance
of an evanescent (kinetic) operator. Such operators only exist in 
$d$-dimensions and when restricted to two dimensions collapse or evaporate to 
zero. However, their presence in a dimensionally regularized renormalization
is fundamental to ensuring the model can be rendered finite and they affect the
determination of the true renormalization group functions in a subtle way. 
Ignoring their presence and effect would mean that agreement between 
perturbative expressions and the Knizhnik-Zamolodchikov results would not be
possible as emphasised in \cite{4,5}. 

Having recalled these connections between the WZWN model and general results 
one natural question which arises is to what extent can one derive or verify
similar results in explicit perturbation theory in the NATM which is connected 
via the bosonization rules of \cite{2}. To appreciate the non-trivial nature of
such a problem it is worth considering a few salient properties of the 
fermionic model. One interesting feature which it possesses is a connection
with quantum chromodynamics, (QCD), in $d$-dimensions. In \cite{6} it was 
argued that the strong coupling limit of the NATM in $d$-dimensions possesses a
gauge symmetry. Moreover, if one examined the resulting effective action in 
$d$-dimensions in the large $\Nf$ limit where $\Nf$ is the number of quark 
fields, then the Feynman rules of QCD emerge naturally. In essence the closed 
quark loops with the requisite number of external points reproduced the Feynman
rules for the triple and quartic gluon vertices in the limit of going to four 
dimensions, \cite{6}. Higher point functions vanished in the same limit. This 
remarkable connection in $d$-dimensions with the NATM was exploited in a 
variety of papers, \cite{7,8,9}, to compute $d$-dimensional critical exponents.
These provided all orders results in QCD perturbation theory in the strong 
coupling constant and agreed with the corresponding explicit high order 
perturbative results in the region of overlap. Moreover, one could go beyond 
the leading order in the large $\Nf$ expansion and compute exponents at 
$O(1/\Nf^2)$, \cite{9}, which agreed with the known four loop $\MSbar$ results
of \cite{10,11}, so that the connection of \cite{6} is confirmed to be not 
solely leading order. Since such a QCD NATM connection is now established one 
can place properties of QCD perturbation theory in the context of the NATM as 
well as the WZWN model. For instance, the four loop $\beta$-function of QCD has
been computed in $\MSbar$ in \cite{12}. One novel feature of this result was 
the appearance of new colour group Casimirs beyond the elementary ones already 
mentioned. They also arise in the quark mass anomalous dimension, \cite{11}, 
and must be present in the quark anomalous dimension at four loops. However, 
whilst this latter quantity has been calculated in QCD in \cite{13}, it was 
only for the colour group $SU(\Nc)$. One cannot explicitly deduce the result 
for an arbitrary Lie group from this result since there is no {\em unique} 
construction of the new general Casimirs from the $\Nc$ dependent information. 
However, by examining the topology of Feynman diagrams it is clear that the new
Casimirs can be present. Therefore, having recalled these connections and group
theory one issue which arises is to do with the structure of the four loop 
anomalous dimension in the NATM. On the one hand the connection with QCD and 
the Feynman diagram topologies indicates that the new Casimirs ought to be 
present. However, the relation of the WZWN model and the Knizhnik-Zamolodchikov
critical exponents suggests that when the renormalization group functions are 
evaluated at criticality either the new Casimir structures somehow cancel or 
they are not present in the first instance. Therefore, this is the main problem
we address in this paper where we will compute the four loop fermion anomalous 
dimension in the NATM to ascertain the group structures which will appear. This
extends the earlier two and three loop calculations of respectively \cite{14} 
and \cite{15}. However, it is far from a straightforward renormalization since 
we will use dimensional regularization and like the treatment of the WZWN model 
the NATM ceases being multiplicatively renormalizable. This was recognised in 
\cite{16,17,18,19} and the necessary formalism was introduced to handle the 
contributions from the new evanescent four fermi operators which result in the 
renormalization based on \cite{20}. In \cite{15} the dimensionally regularized 
two loop $\beta$-function calculation of \cite{14} was extended to three loops 
and the terms beyond that originally computed in the cutoff regularized 
calculation of Destri, \cite{14}, were produced. Whilst the result for a 
general colour group was quoted for the three loop $\beta$-function that 
calculation exploited the fact that to this loop order one could in fact 
compute in the case when the colour group was $SU(\Nc)$ and still construct the
general result uniquely at the end. This was done to speed intermediate aspects
of the computation. As already indicated since such a procedure would fail to 
address the Casimir problem at four loops, we choose here to consider a general 
(classical) Lie group and not specify to an $SU(\Nc)$ group in contrast to 
\cite{15}. Whilst this will lead to a longer calculation it will reveal a rich 
group structure at various stages. Moreover, we will address the evanescent 
operator issue in the general group case and establish the renormalization 
group functions akin to those which underpinned the check of the 
Knizhnik-Zamolodchikov exponents in the WZWN model. Given the nature of the 
NATM being a four fermi interaction our calculation is intimately connected 
with the $O(N)$ Gross Neveu model, \cite{21}, whose four loop anomalous 
dimension is known in $\MSbar$, \cite{22}, which extended the lower order 
calculations of \cite{23,24}. This will provide an important cross-check on our
integration routines which have been implemented in a symbolic manipulation 
package called {\sc Form}, \cite{25}. By ensuring that the correct four loop 
Gross Neveu result is established in our programmes, changing the input 
Lagrangian to the NATM means one can essentially rely on that aspect of our 
calculation being correct, aside from the usual internal checks one usually 
has. Although we will produce the $\MSbar$ result it is important to note that
in the NATM, the Fierz symmetry of two dimensions gets destroyed when one 
computes with a dimensional regularization, \cite{16,17,18,19}. Therefore, we 
will need to consider a finite renormalization beyond minimal subtraction in 
order that the Fierz symmetry and other model equivalences are restored in the 
perturbative renormalization group functions similar to those considered in 
\cite{15}. For example, the single flavour abelian Thirring model is equivalent
to the single flavour Gross Neveu model. However, by examining the three loop 
$\MSbar$ NATM fermion anomalous dimension, \cite{15}, the restriction of that 
result to compare with the same $\MSbar$ result in the Gross Neveu model shows 
that it is not in agreement and therefore in need of a finite renormalization 
to restore its equivalence. Whilst this additional finite renormalization may
at first sight appear to be technical, it is in fact a standard way of dealing
with the limitations of dimensional regularization. In other words it does not
preserve certain symmetries of the theory, such as chiral symmetry, in four 
dimensional gauge theories. The main benefits of using a dimensional 
regularization, which far outweigh other approaches, is the ability to 
calculate to {\em high} loop order.  

The paper is organised as follows. In section 2 we recall the basic properties
of the NATM necessary for the four loop renormalization including the relevant 
lower order renormalization constants. Those for the four fermi evanescent 
operators have been rederived for the case of a general colour group. Section 3
is devoted to the technical details of the full four loop renormalization 
including how certain two loop subgraph Feynman integrals are evaluated by the 
Gram determinant method of \cite{26} and how the evanescent operators are 
handled in the higher order corrections. The results of the full calculation 
are discussed in section 4 where a study of one general set of scheme changes 
that can be performed is given. Our conclusions are given in section 5 and an 
appendix is devoted to the derivation and general aspects of the Gram 
determinant method. 

\sect{Background.} 

The Lagrangian for the strictly two dimensional (massless) non-abelian Thirring
model is  
\begin{equation} 
L^{\mbox{\footnotesize{natm}}} ~=~ i \bar{\psi}^{iI} \partialslash \psi^{iI} 
{}~+~ \frac{g}{2} \left( \bar{\psi}^{iI} \gamma^\mu T^a_{IJ} \psi^{iJ} 
\right)^2 
\label{natmlag} 
\end{equation}  
where $\psi^{iI}$ is a Dirac fermion with flavour and colour indices $i$ and
$I$ respectively with $1$ $\leq$ $i$ $\leq$ $\Nf$ and $1$ $\leq$ $I$ $\leq$ 
$N_{\mbox{\footnotesize{fund}}}$. Here we denote by 
$N_{\mbox{\footnotesize{fund}}}$ the dimension of the fundamental 
representation of the colour group $G$ whose group generators are $T^a$, $1$ 
$\leq$ $a$ $\leq$ $N_A$, and $N_A$ is the dimension of the adjoint 
representation. The generators obey the usual Lie algebra  
\begin{equation} 
[ T^a, T^b ] ~=~ i f^{abc} T^c 
\label{liealg} 
\end{equation} 
where $f^{abc}$ are the structure constants. We have chosen the sign of the 
coupling constant $g$ in a non-standard way. This is in order to make extensive
use of the results of \cite{15} from the point of view of renormalization 
constants but it is elementary to set $g$ $=$ $-$ $\lambda$ where $\lambda$ is
the conventional coupling constant at the end of the calculations to ensure
the model is asymptotically free, \cite{1}. Unlike \cite{15}, however, we will 
not specify the colour group to be $SU(\Nc)$ since we will be concerned with
determining the rich Casimir structure of the anomalous dimension at four 
loops. The advantage of restricting attention to $SU(\Nc)$ in determining the 
three loop $\beta$-function was that the final result could only depend on the
elementary group Casimirs $C_2(R)$ and $C_2(G)$ as well as $T(R)\Nf$ where 
$T^a T^a$ $=$ $C_2(R)$, $f^{acd} f^{bcd}$ $=$ $C_2(G) \delta^{ab}$ and 
$\mbox{Tr}(T^a T^b)$ $=$ $T(R) \delta^{ab}$. Therefore, by working with $T(R)$
$=$ $\half$, $C_2(R)$ $=$ $(\Nc^2-1)/(2\Nc)$ and $C_2(G)$ $=$ $\Nc$ for 
$SU(\Nc)$ it was possible to uniquely determine the result for general $G$ from
the $SU(\Nc)$ value of the $\beta$-function. The potential appearance of new 
Casimirs at four loops means this avenue is closed to us here. Therefore, we 
are forced to work with a general Lie group. For definiteness the (three) new 
Casimirs are products of the symmetric tensors, \cite{12,27}, 
\begin{equation} 
d^{abcd}_F ~=~ \mbox{Tr} \left( T^a T^{(b} T^c T^{d)} \right) ~~~,~~~  
d^{abcd}_A ~=~ \mbox{Tr} \left( A^a A^{(b} A^c A^{d)} \right) ~~~,~~~  
\end{equation} 
where $(A^a)_{bc}$ $=$ $-$ $i f^{abc}$ is the adjoint representation of the
Lie algebra. In particular for $SU(\Nc)$, \cite{12}, 
\begin{equation}  
\frac{d^{abcd}_F d^{abcd}_F}{N_A} ~=~ \frac{(\Nc^4 - 6\Nc^2 + 18)}
{96\Nc^2} ~~~,~~~ \frac{d^{abcd}_A d^{abcd}_F}{N_A} ~=~ 
\frac{\Nc(\Nc^2+6)}{48} ~. 
\end{equation} 

Whilst we have to adopt a different strategy to deal with the group theory of 
(\ref{natmlag}) the approach to compute the anomalous dimension of $\psi^{iI}$
remains the same as \cite{15,16,17,18,19}. In \cite{14} the two loop 
calculation was performed in strictly two dimensions using a momentum cutoff 
and a version of (\ref{natmlag}) which involves an auxiliary spin-$1$ field, 
which in $d$-dimensions would correspond to the gluon of QCD at the large $\Nf$ 
non-trivial fixed point, \cite{6}. Indeed this is the reason for choosing 
$\psi^{iI}$ to be in the fundamental representation of $G$. However, using a 
cutoff to regularize the integrals is not a feasible approach beyond two loops 
despite the advantages that exist from maintaining a $\gamma$-algebra which is 
strictly two dimensional. Instead it is better to use dimensional 
regularization retaining the $\MSbar$ scheme since it allows one to compute the
massless Feynman integrals more easily. The major difficulty in this is the 
loss of the finite dimensional $\gamma$-algebra. See, for example, \cite{28}. 
In two dimensions, for instance, the identity  
\begin{equation} 
\gamma^\mu \gamma^\nu \gamma_\mu ~=~ 0 
\end{equation}  
reduces the structure of the numerator of an integral substantially given the
nature of the interaction of (\ref{natmlag}). In $d$-dimensions one must use 
\begin{equation} 
\gamma^\mu \gamma^\nu \gamma_\mu ~=~ -~ (d-2) \gamma^\nu 
\end{equation}  
so that extra terms will emerge. However, there is a deeper subtlety involved 
and that is that two dimensional four fermi theories cease to be 
multiplicatively renormalizable in $d$-dimensions. For the NATM the breakdown 
is at one loop and was pointed out in \cite{16,17}. Latterly for the Gross 
Neveu model it has been shown in \cite{29} that this model is also not 
multiplicatively renormalizable in dimensional regularization with the first 
occurence being at three loops. This has been verified in \cite{15}. This 
feature is evidenced in the appearance of evanescent operators in the 
renormalization of the four point function. In other words new four point 
operators are generated under the renormalization which do not have the 
structure of the original interaction term and have the evanescent property 
that they do not exist in the two dimensional limit. This may suggest that 
these extra operators play no role in the construction of the strictly two 
dimensional renormalization group functions. However, if we focus on the NATM 
beyond one loop one has to include these new operators in the higher loop 
Green's functions and more importantly account for their effect and 
contribution in the renormalization group functions in $d$-dimensions prior to 
restricting the $\MSbar$ results to two dimensions. We recall that when one 
calculates the $\beta$-function in a theory (without evanescent operators) one 
works with renormalization constants, $Z$, which depend on $\epsilon$ where $d$
$=$ $2$ $-$ $\epsilon$ and a coupling constant which is dimensionless in 
$d$-dimensions. In deducing $\beta(g)$ correctly the dimensionality of the 
coupling constant in $d$-dimensions is crucial in restoring terms which are of 
the form $\epsilon/\epsilon$. Therefore when evanescent operators are present 
they in fact subtlely contribute to the renormalization group function in such 
a way that na\"{\i}vely following the usual procedure described above, their 
presence remains in such renormalization group functions. Hence one would 
obtain incorrect results for the strictly two dimensional functions. Indeed 
this point can best be appreciated if one calculates the two loop 
$\beta$-function of the NATM using the na\"{\i}ve approach. In \cite{17} a 
result emerges for this $\beta$-function which does not agree with the two loop
result of \cite{14}. It is obvious that the $d$-dimensional calculation cannot 
be correct since for a single coupling theory the two loop $\beta$-function is 
scheme independent. 

There are various well established ways of accounting for the effect of the 
evanescent operators in the renormalization group functions. One approach 
developed in \cite{20} and used in the context of two dimensional four fermi 
theories is that of the projection or reduction formula. By examining the 
renormalized Lagrangian with the full set of evanescent operators incorporated 
one can relate any true renormalization group function to the na\"{\i}ve ones 
which have evanescent effects present by adding terms in a well defined 
procedure which exactly accounts for the projected contribution from the 
evanescent operators, \cite{20}. In essence the contribution of the evanescent 
operator inserted in the appropriate Green's function but evaluated after 
operator insertion renormalization in strictly two dimensions, accounts for the
overcounting in the na\"{\i}ve renormalization group functions. The alternative
approach which we will use here and in \cite{15} is to examine the 
renormalization group equation, which involves the true renormalization group 
functions, acting on the finite renormalized Green's function. As this is a 
purely two dimensional object the presence of the evanescent operators is 
effectively washed out and hence one can deduce the renormalization group 
functions by ensuring the renormalization group equation is valid to the order 
in perturbation theory one is interested in. However, for this approach to be 
successful one has to evaluate all the Feynman integrals at that loop order to 
the finite part inclusively. Ordinarily when renormalizing a theory at high 
loop order determining the finite part is a difficult aspect of the 
calculation. However, in using this strategy here it will turn out that it is 
possible to determine all the necessary parts of the four loop Feynman graphs 
to achieve this aim. Moreover, we have checked our final answers explicitly for
the case $G$ $=$ $SU(\Nc)$ by using the projection formula of \cite{20} as an 
independent verification. 

Having reviewed the issue of evanescent operators in the $d$-dimensional 
extension of the NATM in relation to the construction of the renormalization 
group functions, we will now introduce their explicit forms for the 
calculation. First, we need to define the basis for the $\gamma$-algebra in
$d$-dimensions which has been introduced in \cite{16,18,30,31}. In 
$d$-dimensions the Clifford algebra  
\begin{equation} 
\{ \gamma^\mu , \gamma^\nu \} ~=~ 2 \eta^{\mu\nu} 
\end{equation} 
remains central to all $\gamma$-matrix manipulations except that the Lorentz
indices now run over $1$~$\leq$~$\mu$~$\leq$~$d$ where $d$ is non-integer. This
latter property means that, for example, the anti-symmetrized product of five
or more $\gamma$-matrices is non-zero in $d$-dimensions whereas they would
clearly vanish in four dimensions. Therefore, this set of this combination of 
$\gamma$-matrices is infinite dimensional in $d$-dimensions and can be used to 
span the space occupied by the $d$-dimensional $\gamma$-matrices. For 
compactness we introduce the objects $\Gamma_{(n)}^{\mu_1 \ldots \mu_n}$ which 
are anti-symmetric in the Lorentz indices and defined by 
\begin{equation} 
\Gamma^{\mu_1 \ldots \mu_n}_{(n)} ~=~ \gamma^{[\mu_1} \ldots \gamma^{\mu_n]}
\end{equation} 
as the basis for the $\gamma$-matrices when we treat (\ref{natmlag}) in
$d$-dimensions. Before considering a general group $G$ it is instructive to 
recall the structure of evanescent operators for the case $G$ $=$ $SU(\Nc)$ as
it will also illustrate other issues such as the restoration of multiplicative 
renormalizability. For $SU(\Nc)$ the group generators 
satisfy\footnote{Analogous decompositions exist for the other classical and
exceptional groups. See, for example, \cite{32}.}  
\begin{equation} 
T^a_{IJ} T^a_{KL} ~=~ \frac{1}{2} \left[ \delta_{IL} \delta_{KJ} ~-~ 
\frac{1}{\Nc} \delta_{IJ} \delta_{KL} \right] ~.  
\label{suncdeco} 
\end{equation} 
This means that when performing loop calculations based on the interaction of
(\ref{natmlag}) any string of group generators involving an even number of
$T^a$'s can be written in terms of the $SU(\Nc)$ tensor basis of 
(\ref{suncdeco}) which is $I \otimes I$ and $T^a \otimes T^a$. Therefore, one 
can immediately write down the most general Lagrangian in $d$-dimensions which 
contains (\ref{natmlag}) in two dimensions. We have, \cite{15,18,19}, 
\begin{eqnarray} 
L^{\mbox{\footnotesize{natm}}} &=& i \bar{\psi}^{iI} \partialslash \psi^{iI} 
{}~+~ \frac{g}{2} \left( \bar{\psi}^{iI} \gamma^\mu T^a_{IJ} \psi^{iJ} 
\right)^2 ~+~ \frac{1}{2} \sum_{k=0}^\infty g_{k0} \left( \bar{\psi}^{iI} 
\Gamma_{(k)} \delta_{IJ} \psi^{iJ} \right)^2 \nonumber \\ 
&& +~ \frac{1}{2} \sum_{k=0, k \neq 1}^\infty g_{k1} \left( \bar{\psi}^{iI} 
\Gamma_{(k)} T^a_{IJ} \psi^{iJ} \right)^2 
\label{sunclag}
\end{eqnarray}  
where all quantities are bare and the Roman letter $k$ labels evanescent 
contributions in the same notation as \cite{15}. As in \cite{18,19,15} we have 
introduced a coupling constant for each evanescent operator which thereby 
ensures (\ref{sunclag}) is now multiplicatively renormalizable in 
$d$-dimensions. Ordinarily one would use (\ref{sunclag}) to determine the true 
renormalization group functions but from the point of view of calculability 
(\ref{sunclag}) is unpractical. To see this one need only consider the fact 
that at one loop each operator needs to be renormalized and given the nature of
the one loop diagrams there will be mixing between all the operators so that 
proceeding to higher orders becomes a huge exercise. For instance, the vertex 
$(\bar{\psi} T^a \gamma^\mu \psi)^2$ generates the additional vertices 
$(\bar{\psi} I \Gamma_{(3)} \psi)^2$ and $(\bar{\psi} T^a \Gamma_{(3)} \psi)^2$
at one loop, \cite{17,15}. However, to achieve the aim of determining the two 
dimensional renormalization group functions this is not necessary since in the 
limit to two dimensions the new operators will be absent or equivalently one 
can set the new evanescent couplings to zero. In this case new operators will 
be generated under renormalization with a coupling which is the effective 
renormalization constant. As there will be a finite number of these at each new
loop order it is much easier to calculate their effect and renormalization 
compared to an infinite set. Their full presence in the na\"{\i}ve 
renormalization group equations are then accounted for by the formalism of 
\cite{20,17}, discussed earlier. Having recalled the structure for the case $G$
$=$ $SU(\Nc)$ the general group situation is only complicated by the absence of
a general identity (\ref{suncdeco}). In other words the complete set of 
operators analogous to those of (\ref{sunclag}) in $SU(\Nc)$ will still involve
the $\Gamma_{(n)}$-matrices but the group theory content has to be replaced by 
general functions of the group generators which are a basis for the group 
space. Whilst we do not know how to determine this to all orders in general we 
instead define the $d$-dimensional Lagrangian with evanescent four fermi 
operators as  
\begin{equation} 
L^{\mbox{\footnotesize{natm}}} ~=~ i \bar{\psi}^{iI} \partialslash \psi^{iI} 
{}~+~ \frac{g}{2} \left( \bar{\psi}^{iI} \gamma^\mu T^a_{IJ} \psi^{iJ} 
\right)^2 ~+~ \sum_{k=0}^\infty \sum_{\{\alpha\}} \sum_{\{\beta\}} 
g_{k \, \alpha \beta} {\cal O}_k^{\alpha \beta}  
\end{equation}  
where 
\begin{equation} 
{\cal O}_k^{\alpha \beta} ~=~ \frac{1}{2} \left( \bar{\psi}^{iI} 
\Gamma_{(k)} {\cal T}^\alpha_{IJ} \psi^{iJ} \right) \left( \bar{\psi}^{jK} 
\Gamma_{(k)} {\cal T}^\beta_{KL} \psi^{jL} \right) ~.  
\end{equation} 
Here ${\cal T}^\alpha_{IJ}$ are functions of the group generators and the 
indices $\alpha$ and $\beta$ represent sets of free colour group indices $A$. 
If one knew the full basis explicitly it would be possible to relate the 
$SU(\Nc)$ evanescent couplings $g_{k \, \alpha\beta}$ to those of the general 
case. As we will work with nullified evanescent couplings some insight into the
possible form of the group basis can be gained by examining the renormalization
of the four point function, \cite{15}. This is in fact necessary here because 
we need to know which operators are generated as they have to be included in 
the subsequent order of the two point function diagrams and their contribution 
to the na\"{\i}ve wave function renormalization accounted for. In \cite{17} the
new operators generated in the one loop four point function were calculated for 
general $G$ and incorporated into the two loop calculation. However, the new 
operators generated at that order were not recorded. Whilst they were 
determined in \cite{15} at two and three loops for $SU(\Nc)$ these results were
expressed in the $\{ I\otimes I, T^a \otimes T^a \}$ basis which also cannot be
uniquely generalized to arbitrary $G$. Therefore, we have recalculated the full
two loop $4$-point function renormalization for general $G$. This involved 
computing the graphs of figure 1 which generated the evanescent operator  
${\cal O}_{32}$ $=$ $\half \left( \bar{\psi}^{iI} \Gamma_{(3)} T^{(ab)}_{IJ} 
\psi^{iJ} \right)^2$ at one loop where the operator is labelled according to 
the respective number of $\gamma$-matrices and group generators present. In 
\begin{figure}[ht] 
\epsfig{file=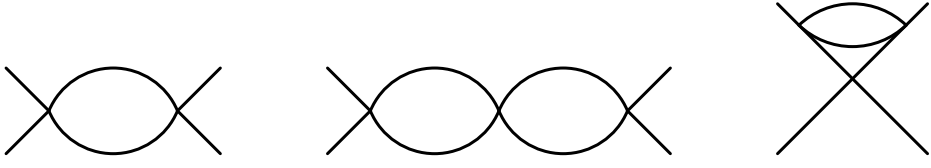,height=2cm} 
\vspace{0.5cm} 
\caption{One and two loop corrections to the $4$-point function.} 
\end{figure} 
figure 1 the vertex corresponds to the original vertex of (\ref{natmlag}) 
whilst for the full two loop calculation the additional graphs of figure 2 must
be included where the symbol $\otimes$ at a vertex there represents the 
insertion of the operator ${\cal O}_{32}$. We find that the full renormalized 
Lagrangian to two loops is  
\begin{eqnarray} 
L^{\mbox{\footnotesize{natm}}} &=& i Z_\psi \bar{\psi}^{iI} \partialslash 
\psi^{iI} ~+~ \frac{g}{2} {\tilde{\mu}}^\epsilon Z_g Z_\psi^2 
\left( \bar{\psi}^{iI} \gamma^{\tilde{\mu}} T^a_{IJ} \psi^{iJ} \right)^2 
\nonumber \\
&& +~ \frac{g}{2} {\tilde{\mu}}^\epsilon Z_{32} Z_\psi^2 \left( \bar{\psi}^{iI} 
\Gamma_{(3)} T^{(ab)}_{IJ} \psi^{iJ} \right)^2 
{}~+~ \frac{g}{2} {\tilde{\mu}}^\epsilon Z_{13} Z_\psi^2 \left( \bar{\psi}^{iI} 
\Gamma_{(1)} T^{(abc)}_{IJ} \psi^{iJ} \right)^2 \nonumber \\  
&& +~ \frac{g}{2} {\tilde{\mu}}^\epsilon Z_{31} Z_\psi^2 \left( \bar{\psi}^{iI} 
\Gamma_{(3)} T^{a}_{IJ} \psi^{iJ} \right)^2 
{}~+~ \frac{g}{2} {\tilde{\mu}}^\epsilon Z_{33} Z_\psi^2 \left( \bar{\psi}^{iI} 
\Gamma_{(3)} T^{[abc]}_{IJ} \psi^{iJ} \right)^2 \nonumber \\  
&& +~ \frac{g}{2} {\tilde{\mu}}^\epsilon Z_{51} Z_\psi^2 \left( \bar{\psi}^{iI} 
\Gamma_{(5)} T^{a}_{IJ} \psi^{iJ} \right)^2 
{}~+~ \frac{g}{2} {\tilde{\mu}}^\epsilon Z_{53} Z_\psi^2 \left( \bar{\psi}^{iI} 
\Gamma_{(5)} T^{(abc)}_{IJ} \psi^{iJ} \right)^2 
\label{genlag} 
\end{eqnarray}  
where $\tilde{\mu}$ is the scale introduced to ensure that the coupling 
constant remains dimensionless in $d$-dimensions\footnote{We use $\tilde{\mu}$
as the renormalization scale throughout instead of the conventional $\mu$ since
the latter will subsequently denote $d/2$.} and we note that the evanescent 
operators all involve the symmetric product of group generators except 
$Z_{33}$. Here the fields and the 
\begin{figure}[hb] 
\epsfig{file=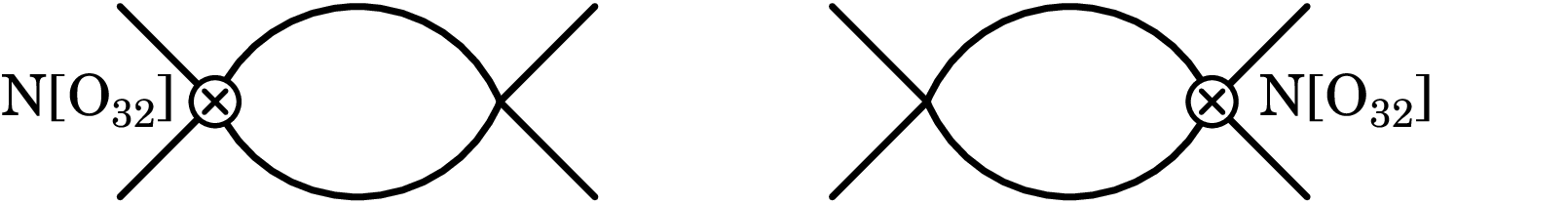,height=1.25cm} 
\vspace{0.5cm} 
\caption{Single evanescent operator contributions to the $2$ loop $4$-point 
function.} 
\end{figure} 
coupling constant $g$ are regarded as the renormalized ones as opposed to those
in (\ref{natmlag}) and (\ref{sunclag}) which were assumed to be bare. We find 
the wave function renormalization constant is 
\begin{eqnarray} 
Z_\psi &=& 1 ~-~ \frac{C_2(R)T(R)\Nf g^2}{4\pi^2\epsilon} ~-~ C_2(R)  
\left[ \frac{C_2(G)T(R)\Nf}{12\epsilon^2} \right. \nonumber \\ 
&& \left. ~~+~ \frac{[4T^2(R)\Nf^2 + C_2(G)T(R)\Nf + 2C^2_2(G) 
- 6C_2(G) C_2(R) + 4C^2_2(R)]}{24\epsilon} \right] \frac{g^3}{\pi^3} ~+~ O(g^4) 
\nonumber \\ 
\end{eqnarray} 
and the four point operator renormalization constants to $O(g^3)$ are 
\begin{eqnarray} 
Z_g &=& 1 ~+~ \frac{C_2(G)g}{2\pi\epsilon} ~+~ \left( \frac{C^2_2(G)} 
{4\epsilon^2} ~-~ \frac{[C^2_2(G) - 8C_2(G) T(R) \Nf]}{32\epsilon} \right) 
\frac{g^2}{\pi^2} \nonumber \\ 
Z_{32} &=& -~ \frac{g}{2\pi\epsilon} ~+~ 
\left[ \frac{[7C_2(G) - 4C_2(R)]}{2\epsilon^2} ~+~ 
\frac{[7C_2(G) - 2C_2(R) - 2 T(R) \Nf]}{4\epsilon} \right] \frac{g^2}{\pi^2}  
\nonumber \\ 
Z_{13} &=& -~ \frac{3g^2}{2\pi^2\epsilon} ~~~,~~~  
Z_{33} ~=~ \left[ \frac{15}{\epsilon^2} ~+~ \frac{15}{2\epsilon} \right] 
\frac{g^2}{\pi^2} ~~~,~~~ 
Z_{51} ~=~ C^2_2(G) \left[ \frac{1}{192\epsilon^2} ~+~ \frac{1}{384\epsilon}
\right] \frac{g^2}{\pi^2} \nonumber \\
Z_{53} &=& \left[ \frac{1}{4\epsilon^2} ~-~ \frac{1}{16\epsilon} \right] 
\frac{g^2}{\pi^2} ~.  
\end{eqnarray}  
These values reduce to those quoted in \cite{15} for the case $G$ $=$ 
$SU(\Nc)$. Although to two loops we have managed to write the generated 
operators in terms of anti-symmetric or symmetric products of strings of group 
generators, it is not clear to us whether this should persist to all orders. 
For example, it may be the case that at a very large loop order the Lie algebra
manipulations we used to obtain this Lagrangian do not allow us to reduce the
generator products to pure $T^a$-strings in that one or more structure 
functions may remain or three or more traces of generators. A systematic study 
which goes beyond the symmetric traces of group generators for general $G$ of 
\cite{27} would be needed. However, to the order we will be computing the wave 
function anomalous dimension, (\ref{genlag}) is all that will be necessary if
one considers the topologies of the diagrams which will arise and the operators 
which can be inserted at various vertices. 

We close this section by recalling the relationship of the NATM with other
models. As is already well established the abelian Thirring model for $\Nf$ $=$
$1$ is trivially equivalent to the single flavour Gross Neveu model due to the
Fierz identity for the Dirac fermions we are using which is  
\begin{equation} 
(\bar{\psi} \gamma^\mu \psi)^2 ~=~ -~ 2 (\bar{\psi} \psi)^2 ~.  
\label{fierz} 
\end{equation} 
Therefore, the renormalization group functions for both models in this limit
must agree. However, when one examines the $\MSbar$ renormalization group 
function in each case it transpires that beyond the leading order there is 
disagreement which arises because (\ref{fierz}), which is established using 
the properties of the strictly two dimensional $\gamma$-algebra, is not valid
in $d$-dimensions. Indeed one can use a $d$-dimensional Fierz lemma, 
\cite{33,34}, to discover that an infinite set of additional operators emerge 
on the right side of (\ref{fierz}) in this case. Therefore, to restore the 
clear equivalence between the renormalization group functions in both models 
one needs to make a finite scheme change in the NATM results. Introducing a 
finite renormalization to ensure a symmetry principle is preserved in the 
quantum theory is not novel to the NATM. For example, the renormalization of 
the flavour singlet axial vector current in QCD in $\MSbar$ ceases to preserve 
the chiral anomaly and one ensures full consistency by a finite 
renormalization, \cite{35,36}. Whilst it may appear here that the problem is 
not related to the $\gamma^5$ issue of the QCD axial anomaly, it is in fact 
implicit in the computation since the $d$-dimensional object 
$\Gamma^{\mu\nu}_{(2)}$ would be the projector of the two dimensional 
$\gamma^5$. Therefore, as in \cite{15} it will be the case that having 
established an $\MSbar$ result for the four loop anomalous dimension we will
need to determine the constraints on a class of possible scheme changes to 
preserve the Gross Neveu equivalence. In \cite{17} another equivalence was
mentioned between the $\Nf$ $=$ $3$ Gross Neveu model and the $\Nf$ $=$ $1$ 
$SU(4)$ NATM. Whilst the implications of this were discussed for the three loop
NATM $\beta$-function, \cite{15}, we will not impose it or discuss it here 
primarily because it is not clear if a full rigorous proof of the relationship 
has been established. If one does exist it will be a trivial exercise to 
determine its consequences for the anomalous dimension. For later use and to 
illustrate these remarks we now quote the relevant renormalization group 
functions in $\MSbar$ in both models to the orders they are known. For the 
Gross Neveu model the anomalous dimension in our notation is 
\cite{22,23,24},
\begin{eqnarray} 
\gamma(g) &=& -~ \frac{(2\Nf-1)g^2}{8\pi^2} ~+~ 
\frac{(2\Nf-1)(\Nf-1)g^3}{16\pi^3} \nonumber \\ 
&& -~ \frac{(2\Nf-1)(4\Nf^2 - 14\Nf + 7)g^4}{128\pi^4} ~+~ O(g^5) 
\end{eqnarray}  
and the $\beta$-function is, \cite{21,23,24,37,38}, 
\begin{equation} 
\beta(g) ~=~ -~ \frac{(\Nf-1)g^2}{\pi} ~+~ \frac{(\Nf-1)g^3}{2\pi^2} ~+~ 
\frac{(2\Nf-7)(\Nf-1)g^4}{16\pi^3} ~+~ O(g^5) \\ 
\end{equation} 
where we note that the Gross Neveu Lagrangian is, \cite{21},  
\begin{equation} 
L^{\mbox{\footnotesize{GN}}} ~=~ i \bar{\psi}^i \partialslash \psi^i 
{}~+~ \frac{g}{2} \left( \bar{\psi}^i \psi^i \right)^2 ~.  
\label{gnlag} 
\end{equation}  
For the NATM the anomalous dimension is, \cite{14,15,18,19}, 
\begin{eqnarray} 
\gamma(g) &=& -~ \frac{\CR T(R) \Nf g^2}{2\pi^2} \nonumber \\ 
&& +~ \left[ 2\CG\CR - C^2_2(G) - 2\CG T(R) \Nf - 8 T^2(R) \Nf^2 \right] 
\frac{\CR g^3}{16\pi^3} ~+~ O(g^4) \nonumber \\  
\end{eqnarray} 
and the $\beta$-function is, \cite{1,14,17,15}, 
\begin{eqnarray} 
\beta(g) &=& \frac{\CG g^2}{2\pi} ~+~ \frac{T(R)\Nf\CG g^3}{2\pi^2} 
\nonumber \\ 
&&+~ \CG \left[ \frac{5}{8} T^2(R) \Nf^2 ~+~ \frac{39}{16} C^2_2(R) ~-~ 
\frac{67}{32}\CR\CG ~+~ \frac{31}{64} C^2_2(G) \right] \frac{g^4}{\pi^3} ~+~ 
O(g^5) ~. \nonumber \\  
\end{eqnarray}  
We also note that several all orders results are available in the Thirring
model, though it is not clear if they are in an $\MSbar$ scheme. In the 
abelian case, \cite{39}, 
\begin{equation}
\gamma(g) ~=~ -~ \frac{\Nf g^2}{2\pi^2 \left(1 - \frac{\Nf g}{\pi} 
\right)} ~~~,~~~ \beta(g) ~=~ 0 ~.
\label{abtmrge} 
\end{equation}
For the NATM, Kutasov has written down an expression for $\beta(g)$ which has
a similar form to (\ref{abtmrge}), \cite{40}. In the current notation of this 
paper it is 
\begin{equation}
\beta(g) ~=~ \frac{\CG g^2}{2\pi \left(1 - \frac{T(R) \Nf g}{2\pi} \right)^2}~. 
\label{kutbeta} 
\end{equation}
This expression was established through a connection the two dimensional theory
has with string and conformal field theory but was in fact only valid when 
$\Nf$ is large. More recently, however, by using current algebra Ward 
identities this result has been argued to be exact at all orders in a 
particular scheme, \cite{41}. This is an important point since it only contains
the elementary Casimirs of $G$ and not the higher order ones which we are 
concerned with here. Whilst (\ref{kutbeta}) would put additional constraints on
potential scheme changes needed for the $\beta$-function we will not need to
consider these here as they will not play a role in the analysis of the 
anomalous dimension. 

\sect{Four loop calculations.} 

We are now in a position to carry out the four loop renormalization of the wave
function in (\ref{natmlag}) using the $d$-dimensional evanescent Lagrangian of 
(\ref{genlag}). First, we redo the three loop calculation of \cite{15,17,18,19}
but for the case of a general Lie group. Unlike \cite{15} we will use the 
massless propagator 
\begin{equation} 
\frac{i\pslash}{p^2} 
\label{propdef} 
\end{equation} 
for the fermion field. In \cite{15} the massive NATM was renormalized to three
loops using a fully massive propagator and a propagator which was massless but
which included a mass term in the denominator to act as an infrared regulator.
This was necessary since that paper considered the renormalization of the full
model including the four point function and not just the two point function as
we are here. For the former Green's function such an infrared regulator is 
necessary to avoid potential spurious infrared infinities which could arise 
when using completely massless propagators with two of the external momenta
nullified. In the presence of a mass with all external momenta nullified means 
that the poles in $\epsilon$ which emerged represented ultraviolet infinities
only. For the two point function in the fermionic theory we are considering 
such spurious infrared infinities ought not to arise at the order we are 
working to since, for instance, neither external momentum is nullified. 
Therefore, in all our calculations we use (\ref{propdef}) without an infrared
mass regularization and this allows us in fact to calculate all the necessary 
integrals exactly to four loops or at least to the finite part in $\epsilon$. 
This latter property is important since we will be considering scheme changes 
later and it is necessary to have the renormalized Green's function in order to
extract the renormalization group function in various schemes. Further, by  
choosing a massless propagator the Feynman diagrams with tadpoles\footnote{By 
tadpole in a theory with a four point interaction we mean diagrams with zero 
momentum insertions.} are absent since by Lorentz invariance the tadpole 
subintegral is zero. Therefore to three loops the only two basic topologies 
which are relevant for the wave function renormalization are those given in 
figure 3. They can be evaluated exactly as a function of $d$ $=$ $2\mu$ 
\begin{figure}[ht]  
\epsfig{file=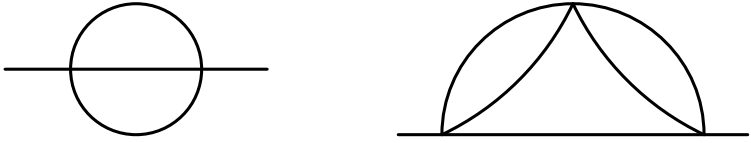,height=1.75cm} 
\vspace{0.5cm} 
\caption{Two and three loop corrections to the $2$-point function.} 
\end{figure} 
and we find that they are respectively  
\begin{equation} 
-~ \frac{ i C_2(R) [(2C_2(R) - C_2(G))(\mu - 1) + 2T(R) \Nf] 
\Gamma(3-2\mu)\Gamma^2(\mu)\Gamma(\mu+1) g^2}{(\mu - 1)\Gamma(3\mu-1)}  
\end{equation} 
and  
\begin{eqnarray} 
&& i \left[ 8(3\mu^2-9\mu+4)C_2(R)C_2(G) ~-~ \frac{16}{3} (4\mu^2-11\mu+5) 
C_2^2(R) ~-~ 16(\mu-1) C_2(R) T(R) \Nf \right. \nonumber \\
&& \left. ~~~-~ \frac{4}{3} (5\mu^2-16\mu+7)C_2^2(G) ~+~ 
\frac{4(6\mu^2-15\mu+7)}{3(\mu-1)} C_2(G) T(R) \Nf ~-~ \frac{16}{3} 
T^2(R) \Nf^2 \right] \nonumber \\ 
&& ~~ \times \frac{\Gamma(4-3\mu)\Gamma^2(2-\mu)\Gamma(3\mu-2)\Gamma^5(\mu) 
C_2(R) g^3}{(\mu-1)\Gamma(3-2\mu)\Gamma(4\mu-2)\Gamma^2(2\mu-1)} 
\end{eqnarray} 
where we have omitted here and in later values of Feynman graphs, the overall 
power of the momentum and renormalization scale $\tilde{\mu}$, which can be 
restored on dimensional grounds, $\pslash$, where $p$ is the external momentum,
and factors of $(4\pi)^{d/2}$. Also the coupling constant $g$ which will appear
in the values is bare. However, for the full three loop anomalous dimension to 
be deduced correctly we need to take account of the contribution from the 
generated evanescent operator ${\cal O}_{32}$ which is represented by the 
Feynman diagrams of figure 4. In these graphs the $\gamma$-algebra now involves
the $\Gamma_{(3)}$-matrix  
\begin{figure}[ht]  
\epsfig{file=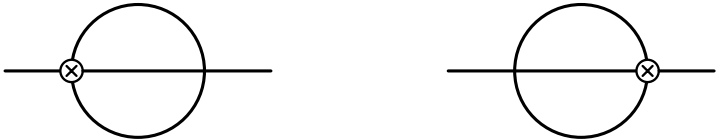,height=1.75cm} 
\vspace{0.5cm} 
\caption{Single evanescent operator contributions to the $3$ loop $2$-point 
function.} 
\end{figure} 
sandwiched between ordinary $\gamma$-matrices. To handle this in the 
computation we decompose $\Gamma_{(3)}$ into its six terms, multiply the 
$\gamma$-string by $\pslash$ and take the trace as well as dividing by the 
normalization of $2p^2$. To cope with the algebra generated and the mundane 
massless integration in this and other graphs we have made use of the symbolic 
manipulation programme {\sc Form}, \cite{25}. Moreover, to ensure that the 
integration routines are properly constructed we have repeated the full four 
loop anomalous dimension calculation of \cite{22} in the $SU(N)$ Gross Neveu 
model and reproduced the result of \cite{22}. Having verified the programmes 
reproduce the correct results for this model it is then elementary to re-run 
them where the input Lagrangian is replaced by the NATM. Whilst evanescent 
operators will not be present in the four loop Gross Neveu calculation the 
integration routines were still valid for the NATM since we evaluated 
systematically all the basic scalar integrals which could arise in any of the 
topologies. Although the decomposition of $\Gamma_{(3)}$ for the graph of 
figure 3 was relatively efficient, for higher order $\Gamma_{(n)}$'s or for 
diagrams with more than one evanescent operator insertion, it was much 
more appropriate to make use of the general properties of the $\Gamma_{(n)}$'s 
in these cases. For example, the lemma, \cite{17,18,31},    
\begin{equation} 
\Gamma^{\mu_1 \ldots \mu_n}_{(n)} \Gamma^{\nu_1 \ldots \nu_m}_{(m)} 
\Gamma_{\mu_1 \ldots \mu_n}^{(n)} ~=~ f(n,m) \Gamma^{\nu_1 \ldots \nu_m}_{(m)} 
\end{equation} 
where 
\begin{equation} 
f(n,m) ~=~ (-1)^{nm} (-1)^{n(n-1)/2} \, \left. \frac{\partial^n}{\partial u^n} 
\left[ (1+u)^{d-m} (1-u)^m \right] \right|_{u = 0} 
\end{equation}  
led to the other graphs being determined relatively quickly. Hence, we record
the value of the graphs of figure 4 are 
\begin{equation} 
\frac{i\CR [3\CG - 4\CR][\CG - 2\CR] (2\mu-1)(\mu-3)\mu\Gamma(3-2\mu) 
\Gamma^3(\mu) g^3}{\Gamma(3\mu-1)} ~.  
\label{z32} 
\end{equation} 

We now turn to the details of the four loop calculation. The basic topologies
are given in figure 5 and fall into two classes. The upper two Feynman
\begin{figure}[hb]  
\epsfig{file=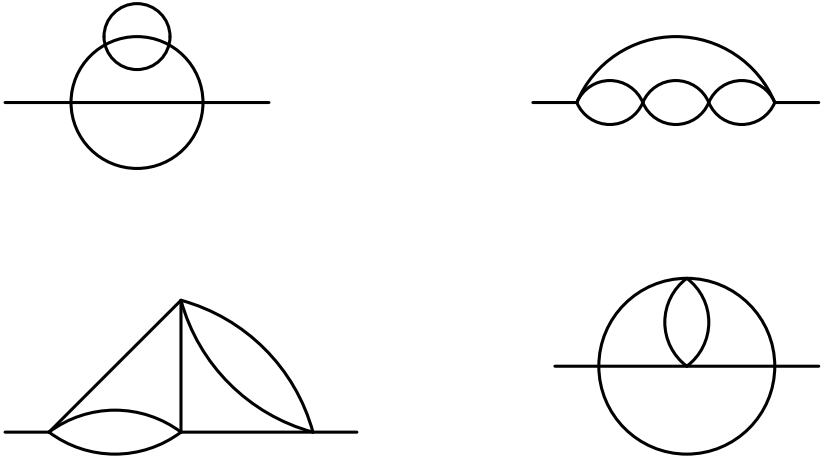,height=6cm} 
\vspace{0.5cm} 
\caption{Four loop corrections to the $2$-point function.} 
\end{figure} 
diagrams correspond to elementary massless chain integrals and can therefore be
computed straightforwardly. We find their $d$-dimensional values are 
respectively 
\begin{eqnarray} 
&& 3 i C^2_2(R) \left[ 2 C_2(G) C_2(R) ~-~ 2 C_2^2(R) ~-~ \frac{4 C_2(R) T(R) 
\Nf}{(\mu - 1)} ~-~ \frac{1}{2} C_2^2(G) \right. \nonumber \\ 
&& \left. ~~~~~~~~~~~~+~ \frac{2 C_2(G) T(R) \Nf}{(\mu - 1)} ~-~ 
\frac{2 T^2(R) \Nf^2}{(\mu - 1)^2} \right] 
\frac{\mu^2\Gamma(5-4\mu)\Gamma^5(\mu) g^4}{(3\mu-2)\Gamma(5\mu-3)} 
\end{eqnarray} 
and 
\begin{eqnarray} 
&& i \left[ 
\frac{12(32\mu^4 - 213\mu^3 + 581\mu^2 - 514\mu + 144) C_2(G) C^3_2(R)}
{(\mu - 1)} \right. \nonumber \\ 
&& \left. ~~~-~ \frac{4(56\mu^4 - 301\mu^3 + 701\mu^2 - 586\mu + 160) C^4_2(R)} 
{(\mu - 1)} \right. \nonumber \\ 
&& \left. ~~~-~ 96(3\mu - 2)(\mu - 1) C^3_2(R) T(R)\Nf ~+~ 
32(3\mu - 2) C_2(G) C_2(R) T^2(R) \Nf^2 \right. \nonumber \\ 
&& \left. ~~~-~ \frac{(232\mu^4 - 1903\mu^3 + 5791\mu^2 - 5294\mu + 1504) 
C^2_2(G) C^2_2(R)}{(\mu - 1)} \right. \nonumber \\ 
&& \left. ~~~+~ 96(3\mu - 2)(\mu - 1) C_2(G) C^2_2(R) T(R) \Nf ~-~ 
\frac{16(3\mu - 2) C_2(R) T^3(R) \Nf^3}{(\mu - 1)} \right. \nonumber \\ 
&& \left. ~~~+~ \frac{(296\mu^4 - 2903\mu^3 + 9479\mu^2 - 8830\mu + 2528) 
C^3_2(G) C_2(R)}{6(\mu - 1)} \right. \nonumber \\ 
&& \left. ~~~-~ \frac{(432\mu^5 - 2004\mu^4 + 3863\mu^3 - 3693\mu^2 + 1734\mu 
- 320) C^2_2(G) C_2(R) T(R)\Nf}{6(\mu - 1)^3} \right. \nonumber \\
&& \left. ~~~-~ \frac{4(8\mu^4 - 125\mu^3 + 461\mu^2 - 442\mu + 128) d_A^{abcd} 
d_F^{abcd}}{(\mu - 1) N_{\mbox{\footnotesize{fund}}}} \right. \nonumber \\
&& \left. ~~~-~ \frac{2(12\mu^4 + 119\mu^3 - 237\mu^2 + 150\mu - 32) d_F^{abcd} 
d_F^{abcd} \Nf}{(\mu - 1)^3 N_{\mbox{\footnotesize{fund}}}} \right. 
\nonumber \\ 
&& \left. ~~~-~ 64(3\mu - 2) C^2_2(R) T^2(R)\Nf^2 \right] 
\frac{\Gamma(5-4\mu)\Gamma(4\mu-3)\Gamma^3(2-\mu)\Gamma^7(\mu) g^4} 
{(2\mu-1)^2\Gamma(5\mu-3)\Gamma(4-3\mu)\Gamma^3(2\mu-1)} ~.  
\end{eqnarray} 
Although these are elementary to derive we draw attention to the group theory
structure of each result. The former since it has a propagator correction 
involves the usual products of Casimirs which arise at lower order. By contrast
the embedded chain diagram involves the two new Casimirs constructed from the
tensors $d_F$ and $d_A$. It is instructive to consider how these arise. Clearly
the term $d_F^{abcd} d_F^{abcd}$ arises from two strings of group generators
where one of the $T^a$ from $T^a \otimes T^a$ tensor structure is located in 
one string with the other in the second string. On the other hand the second
new combination arises from manipulating particular strings of eight
generators. They are all related to  
$\mbox{Tr} \left( T^a T^b T^c T^d T^a T^b T^c T^d \right)$. Using the Lie 
algebra, (\ref{liealg}), it is elementary to deduce  
\begin{eqnarray} 
\mbox{Tr} \left( T^a T^b T^c T^d T^a T^b T^c T^d \right) &=&  
\left[ \prod_{r=0}^3 \left( C_2(R) - \frac{r}{2} C_2(G) \right) ~-~ \frac{1}{8}
C_2^3(G) C_2(R) \right] N_{\mbox{\footnotesize{fund}}} \nonumber \\
&& +~ f^{apq} f^{bqr} f^{crs} f^{dsp} \mbox{Tr} \left( T^a T^b T^c T^d 
\right) ~.  
\end{eqnarray}  
Examining the last term reveals that the combination of structure constants is 
equivalent to $\mbox{Tr} \left( A^a A^b A^c A^d \right)$ where $A^a$ is the 
adjoint representation of the generators and hence this string can be related 
to the $d_F^{abcd} d_A^{abcd}$ term. One can also consider the group theory of 
this diagram from the point of view of the connection with QCD. In 
(\ref{natmlag}) it is possible to write the four point interaction in terms of 
a spin-$1$ auxiliary field which in $d$-dimensions is related to the gluon of 
QCD. In such a reformulation the second diagram of figure 5 would correspond to
the first graph of figure 6 where the spring line
\begin{figure} 
\epsfig{file=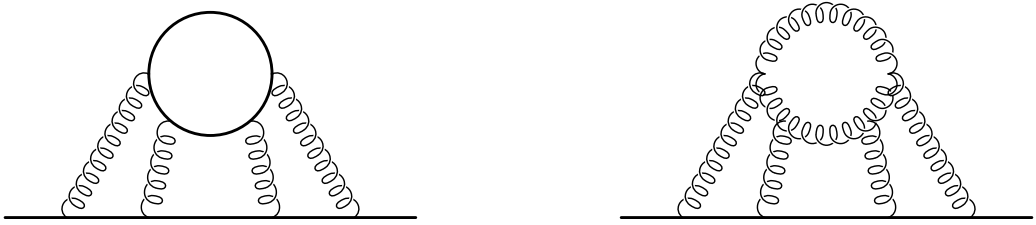,height=3cm} 
\vspace{0.5cm} 
\caption{Four loop self-energy corrections in QCD containing $d_F^{abcd} 
d_F^{abcd}$ and $d_F^{abcd} d_A^{abcd}$ respectively.} 
\end{figure} 
is the auxiliary field. Clearly one can identify the $d_F^{abcd} d_F^{abcd}$ 
structure that is contained in the topology. However, it is not evident how the
$d_F^{abcd} d_A^{abcd}$ structure would correspond. For this case one would
have to have a diagram akin to the second graph of figure 6 where the gluonic 
loop would lead to $\mbox{Tr} \left( A^a A^b A^c A^d \right)$. As this diagram 
cannot be reduced to any of figure 5 by restoring the original interaction it 
might appear that there is a potential inconsistency in our evaluation. 
However, as will become apparent later this issue will be satisfactorily 
resolved. For the remaining two diagrams of figure 5 we expect $d_F^{abcd} 
d_F^{abcd}$ and $d_F^{abcd} d_A^{abcd}$ terms to be present after considering 
the routing of the group generator strings in the graphs. 

The main point concerning these two diagrams is that their evaluation rests in 
the fact that unlike the previous graphs discussed so far they do not reduce
to simple chain integrals. Instead upon taking the spinor trace and evaluating
two elementary loop integrations one is left with a set of two loop integrals
which need to be evaluated. These have the general form of figure 7 where we
\begin{figure}[h]  
\epsfig{file=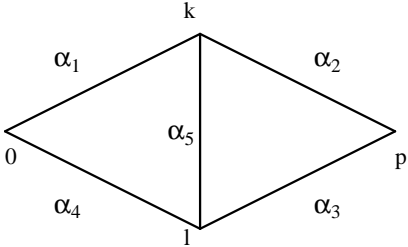,height=3.5cm} 
\vspace{0.5cm} 
\caption{General two loop self energy topology whose value is denoted by 
$\langle \alpha_1, \alpha_2, \alpha_3, \alpha_4, \alpha_5 \rangle$.} 
\end{figure} 
use coordinate space representation where the internal momenta $k$ and $l$ are
integrated over. The $\alpha_i$ beside a line denotes the power to which that
scalar propagator is raised to. The graphs either involve tensor numerators or
are purely scalar. For the former case given the large symmetry of the diagram
one can always rewrite such numerator scalar products in terms of denominator
factors which then reduce the exponent of that line by unity. After this 
procedure one is left with diagrams which are either chains or purely scalar
two loop graphs. For each of the graphs of figure 5 this remaining topology has
the same pattern of exponents. For the third graph we find $\{\alpha_i\}$ $=$
$\{1,\alpha,1,\beta,1\}$ and for the final graph $\{\alpha_i\}$ $=$ 
$\{1,1,1,1,\alpha\}$ where $\alpha$ and $\beta$ here are functions of $d$ and
therefore non-integer when $d$ $=$ $2$ $-$ $\epsilon$. By integration by parts
rules and recurrence relations based on the uniqueness rule of \cite{42,43} one
can reduce this set of graphs either to chain diagrams or to two respective
$d$-dimensional basis scalar integrals given by $\langle 1,1-\mu,1,1-\mu,1 
\rangle$ and $\langle 1,1,1,1,1-\mu \rangle$ where we have again set $d$ $=$ 
$2\mu$ for shorthand. These then are the outstanding diagrams which need to be 
determined. Given the form of the integrals it turns out that they cannot be 
evaluated in closed form for arbitrary values of $d$. However, their $\epsilon$ 
expansion can be deduced to an order beyond that which we need for this 
calculation. This is achieved by using the Gram determinant method of \cite{26}
where the value of a $d$-dimensional two loop self energy diagram of the form 
of that of figure 7 can be {\em exactly} related to a $(d+2)$-dimensional 
diagram of the same form plus a set of elementary chain integrals. In \cite{26} 
the Gram method was applied to a particular choice of the $\{\alpha_i\}$. In 
the appendix we briefly recall the arguments and record the general expression 
for the relation. In, for example, the application of the rule of (\ref{gram}) 
to the set of the form $\langle 1,1,1,1,\alpha \rangle$ the ninth and tenth 
terms of (\ref{gram}) correspond to the same original topology. The actual 
combination of exponents can be restored by the recurrence relations of 
\cite{44}. Having related these two basic integrals to ones in 
$(4-\epsilon)$-dimensions, the $\epsilon$ expansion of the higher dimensional 
integral is determined from the results of \cite{45,46}. Whilst the 
$\epsilon$-expansion is known for the general integral to $O(\epsilon^{4})$, 
\cite{45}, only the first term is relevant to our two dimensional calculation 
as the factors of $(\mu-1)$ ensure the leading term appears in the finite part.
It might be thought that the group theory and symmetry arguments used in 
\cite{45,46} to deduce the $\epsilon$-expansion of the problematic two 
dimensional integrals could be used directly on them. However, these arguments 
seem to rely in part on the fact that in four dimensions the diagram is finite 
with respect to $\epsilon$ and the leading term of the expansion is independent
of $\{\alpha_i\}$ for $\{\alpha_i\}$ of $O(\epsilon)$ from unity. In two 
dimensions the diagram is singular in $\epsilon$ and as the residue of the pole
depends on the parameters it does not appear possible for the group theory 
arguments of \cite{45,46} to proceed. Given these remarks we record the value 
of the third graph of figure 5 is  
\begin{eqnarray} 
&& i \left[ \left( 8(49\mu^4 - 336\mu^3 + 794\mu^2 - 832\mu + 283) C_2(G)
C^3_2(R) \right. \right. \nonumber \\
&& \left. \left. ~~~~-~ 8(27\mu^4 - 174\mu^3 + 392\mu^2 - 394\mu + 131) 
C^4_2(R) \right. \right. \nonumber \\ 
&& \left. \left. ~~~~-~ 8(\mu-1)(18\mu^2 - 59\mu + 32) C^3_2(R) T(R) \Nf ~-~ 
4(6\mu^2 - 15\mu + 8) C^2_2(R) T^2(R) \Nf^2 \right. \right. \nonumber \\ 
&& \left. \left. ~~~~-~ 4(59\mu^4 - 430\mu^3 + 1072\mu^2 - 1178\mu + 411)
C^2_2(G) C^2_2(R) \right. \right. 
\nonumber \\ 
&& \left. \left. ~~~~+~ 4(\mu-1)(42\mu^2 - 161\mu + 88) C_2(G) C^2_2(R) T(R)\Nf
\right. \right. \nonumber \\ 
&& \left. \left. ~~~~+~ \frac{4}{3} (36\mu^4 - 281\mu^3 + 747\mu^2 - 867\mu 
+ 311) C^3_2(G) C_2(R) \right. \right. \nonumber \\
&& \left. \left. ~~~~-~ \frac{(576\mu^4 - 3604\mu^3 + 6853\mu^2 - 5159\mu 
+ 1346) C^2_2(G) C_2(R) T(R) \Nf}{12(\mu - 1)} 
\right. \right. \nonumber \\ 
&& \left. \left. ~~~~+~ 2(6\mu^2 - 29\mu + 16) C_2(G) C_2(R) T^2(R)\Nf^2
\right. \right. \nonumber \\
&& \left. \left. ~~~~-~ \frac{8(3\mu^4 - 38\mu^3 + 144\mu^2 - 210\mu + 83) 
d_A^{abcd} d_F^{abcd}}{
N_{\mbox{\footnotesize{fund}}}} \right. \right. \nonumber \\ 
&& \left. \left. ~~~~-~ \frac{2(4\mu^3 - 37\mu^2 + 23\mu - 2) d_F^{abcd} 
d_F^{abcd} \Nf}{(\mu - 1) N_{\mbox{\footnotesize{fund}}}} \right) 
\frac{\Gamma^2(3-2\mu)\Gamma^6(\mu)}{(\mu-1)^2(2\mu-1)(5\mu-4)\Gamma^2(3\mu-2)}
\right. \nonumber \\ 
&& \left. +~ \left( 384(77\mu^5 - 605\mu^4 + 1743\mu^3 - 2395\mu^2 + 1498\mu 
- 346)(\mu - 1) C_2(G) C^3_2(R) \right. \right. \nonumber \\ 
&& \left. \left. ~~~~~-~ 384(41\mu^5 - 304\mu^4 + 842\mu^3 - 1124\mu^2 
+ 691\mu - 158)(\mu - 1) C^4_2(R) \right. \right. \nonumber \\ 
&& \left. \left. ~~~~~-~ 768(12\mu^3 - 47\mu^2 + 49\mu - 16)(\mu - 1)^2 
C^3_2(R) T(R) \Nf \right. \right. \nonumber \\ 
&& \left. \left. ~~~~~-~ 192(97\mu^5 - 804\mu^4 + 2414\mu^3 - 3424\mu^2 
+ 2183\mu - 510)(\mu - 1) C^2_2(G) C^2_2(R) \right. \right. \nonumber \\ 
&& \left. \left. ~~~~~+~ 1536(7\mu^3 - 32\mu^2 + 34\mu - 11)(\mu - 1)^2 C_2(G) 
C^2_2(R) T(R) \Nf \right. \right. \nonumber \\ 
&& \left. \left. ~~~~~-~ 1536(\mu - 1)^4 C^2_2(R) T^2(R) \Nf^2 \right. \right. 
\nonumber \\ 
&& \left. \left. ~~~~~+~ 32(126\mu^5 - 1103\mu^4 + 3467\mu^3 - 5093\mu^2 
+ 3315\mu - 784)(\mu - 1) C^3_2(G) C_2(R) \right. \right. \nonumber \\ 
&& \left. \left. ~~~~~-~ 8(384\mu^5 - 2713\mu^4 + 6378\mu^3 - 6806\mu^2 
+ 3431\mu - 670) C^2_2(G) C_2(R) T(R) \Nf \right. \right. \nonumber \\ 
&& \left. \left. ~~~~~+~ 192(4\mu^3 - 23\mu^2 + 25\mu - 8)(\mu - 1) C_2(G)
C_2(R) T^2(R) \Nf^2 \right. \right. \nonumber \\ 
&& \left. \left. ~~~~~-~ 384(9\mu^5 - 100\mu^4 + 382\mu^3 - 640\mu^2 + 447\mu 
- 110)(\mu - 1) \frac{d_A^{abcd} d_F^{abcd}}{N_{\mbox{\footnotesize{fund}}}} 
\right. \right. \nonumber \\ 
&& \left. \left. ~~~~~-~ 192(\mu^4 - 18\mu^3 + 14\mu^2 + \mu - 2)
\frac{d_F^{abcd} d_F^{abcd} \Nf}{ N_{\mbox{\footnotesize{fund}}}}
\right) \frac{\Gamma^2(2-\mu)\Gamma^4(\mu)}{6(2\mu-1)(5\mu-4)\Gamma^2(2\mu-1)}
\right. \nonumber \\ 
&& \left. ~~~~~~~\times \left. \frac{}{} \langle 1, 2-\mu , 1, 2-\mu, 1 \rangle 
\right|_{\mu+1} \right. \nonumber \\  
&& +~ \left. \left( \frac{4(19\mu^4 - 126\mu^3 + 294\mu^2 - 304\mu + 105) 
C^4_2(R)}{(\mu - 1)} ~+~ 16(3\mu^2 - 10\mu + 6) C^3_2(R) T(R) \Nf \right. 
\right. \nonumber \\ 
&& \left. \left. ~~~~~~-~ \frac{4(35\mu^4 - 247\mu^3 + 603\mu^2 - 647\mu + 228)
C_2(G) C^3_2(R)}{(\mu - 1)} \right. \right. \nonumber \\ 
&& \left. \left. ~~~~~~+~ \frac{2(43\mu^4 - 322\mu^3 + 826\mu^2 - 924\mu + 333)
C^2_2(G) C^2_2(R)}{(\mu - 1)} \right. \right. \nonumber \\ 
&& \left. \left. ~~~~~~-~ 4(14\mu^2 - 55\mu + 33) C_2(G) C^2_2(R) T(R) \Nf ~+~ 
4(2\mu - 3) C^2_2(R) T^2(R) \Nf^2 \right. \right. \nonumber \\ 
&& \left. \left. ~~~~~~-~ \frac{(54\mu^4 - 431\mu^3 + 1171\mu^2 - 1373\mu + 507)
C^3_2(G) C_2(R)}{3(\mu - 1)} \right. \right. \nonumber \\ 
&& \left. \left. ~~~~~~+~ \frac{(192\mu^4 - 1225\mu^3 + 2387\mu^2 - 1854\mu 
+ 504) C^2_2(G) C_2(R) T(R) \Nf}{12(\mu - 1)^2} \right. \right. \nonumber \\
&& \left. \left. ~~~~~~-~ \frac{4(\mu^2 - 5\mu + 3) C_2(G) C_2(R) T^2(R) 
\Nf^2}{(\mu - 1)} \right. \right. \nonumber \\ 
&& \left. \left. ~~~~~~+~ \frac{4(3\mu^4 - 34\mu^3 + 122\mu^2 - 172\mu + 69)
d_A^{abcd} d_F^{abcd}}{(\mu - 1) N_{\mbox{\footnotesize{fund}}}} \right. 
\right. \nonumber \\ 
&& \left. \left. ~~~~~~+~ \frac{2\mu(\mu^2 - 11\mu + 6) d_F^{abcd} d_F^{abcd} 
\Nf}{(\mu - 1)^2 N_{\mbox{\footnotesize{fund}}}} \right) 
\frac{\Gamma(5-4\mu)\Gamma^5(\mu)}{(\mu-1)(2\mu-1)\Gamma(5\mu-3)} \right] 
g^4 ~. 
\label{nawave41} 
\end{eqnarray}  
The term $\left. \frac{}{} \langle 1, 2-\mu , 1, 2-\mu, 1 \rangle 
\right|_{\mu+1}$ corresponds to the higher dimensional integral derived in the
Gram determinant method where the restriction denotes a $2(\mu+1)$-dimensional
measure. Space prevents us from recording the exact value of the remaining 
graph since it is of a comparable size to (\ref{nawave41}) though we note it 
has a similar form. 

To complete the full four loop calculation the contributions from the 
evanescent operators need to be included. One can divide these into the 
diagrams which involve the operator generated at one loop in the four point
function and those arising from the two loop ones. For the latter, the relevant
topologies are those of figure 2. We record the respective values of the graphs
and include the appropriate renormalization constant to aid identification. 
They are, in addition to (\ref{z32}) which involves $Z_{32}$,  
\begin{eqnarray} 
&& i Z_{13} \left[ 9C_2(G) C^3_2(R) ~-~ 4 C^4_2(R) ~-~ \frac{20}{3} C^2_2(G) 
C^2_2(R) ~+~ \frac{5}{3} C^3_2(G) C_2(R) \right. \nonumber \\  
&& \left. ~~~~~~~-~ \frac{2d_A^{abcd} d_F^{abcd}}
{N_{\mbox{\footnotesize{fund}}}} ~-~ \frac{4d_F^{abcd} d_F^{abcd} \Nf}
{(\mu - 1) N_{\mbox{\footnotesize{fund}}}} \right] 
\frac{\mu\Gamma(3-2\mu)\Gamma^3(\mu)g^2}{\Gamma(3\mu-1)} 
\end{eqnarray} 
\begin{equation} 
4i Z_{31} C_2(R) [2C_2(R) - C_2(G)] \frac{\mu(2\mu-1)(\mu-3)\Gamma(3-2\mu) 
\Gamma^3(\mu)g^2}{\Gamma(3\mu-1)} 
\end{equation} 
\begin{eqnarray} 
&& i Z_{33} \left[ \frac{2}{3} C_2(G) C^3_2(R) ~-~ \frac{2}{3} C^2_2(G) 
C^2_2(R) ~+~ \frac{1}{9} C^3_2(G) C_2(R) \right. \nonumber \\ 
&& \left. ~~~~~~~+~ \frac{4 d_A^{abcd} d_F^{abcd}} 
{3N_{\mbox{\footnotesize{fund}}}} \right] \frac{\mu(2\mu-1)(\mu-3) 
\Gamma(3-2\mu)\Gamma^3(\mu)g^2}{\Gamma(3\mu-1)} 
\end{eqnarray} 
\begin{equation} 
8i Z_{51} C_2(R) [2C_2(R) - C_2(G)] \mu(2\mu - 1)(\mu - 2)(\mu - 5) 
\frac{\Gamma(4-2\mu)\Gamma^3(\mu)g^2}{\Gamma(3\mu-1)} 
\end{equation} 
and 
\begin{eqnarray} 
&& i Z_{53} \left[ 18 C_2(G) C^3_2(R) ~-~ 8 C^4_2(R) ~-~ 
\frac{40}{3}C^2_2(G)C^2_2(R) ~+~ \frac{10}{3} C^3_2(G) C_2(R) \right. 
\nonumber \\ 
&& \left. ~~~~~~~-~ \frac{4 d_A^{abcd} d_F^{abcd}}{ N_{\mbox{\footnotesize{fund}}}} 
\right] \frac{\mu(2\mu-1)(\mu - 5)\Gamma(5-2\mu)\Gamma^3(\mu) g^2} 
{\Gamma(3\mu-1)} ~.  
\end{eqnarray} 
These explicit values illustrate that even though they are two loop topologies 
some do contain the new Casimir structures that appear at four loops. This is 
because unlike the original interaction these operator insertions have more 
\begin{figure}[ht]  
\epsfig{file=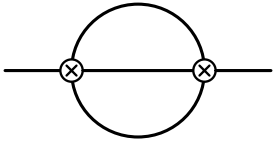,height=1.75cm} 
\vspace{0.5cm} 
\caption{Double evanescent operator correction to the $4$ loop $2$-point 
function.} 
\end{figure} 
than one pair of group generators. Hence the strings of group generators can 
contain one set of eight $T^a$'s. The remaining set of contributions come from 
the insertion of ${\cal O}_{32}$ in the topologies illustrated in figures 8 and
\begin{figure}[h] 
\epsfig{file=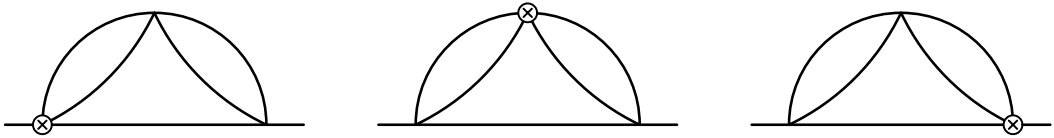,height=1.75cm} 
\vspace{0.5cm} 
\caption{Single evanescent operator correction to the $4$ loop $2$-point 
function.} 
\end{figure} 
9. For the three loop topologies their values are 
\begin{eqnarray}  
&& i Z_{32} \left[ \frac{32(4\mu - 7)(\mu - 3)^2 C^4_2(R)}{(\mu - 1)} ~-~ 
\frac{8(34\mu^2 - 181\mu + 205)(\mu - 3) C_2(G) C^3_2(R)}{(\mu - 1)} 
\right. \nonumber \\ 
&& \left. ~~~~~~~+~ 32(\mu - 3) C^3_2(R) T(R) \Nf ~+~ 
\frac{4(49\mu^2 - 287\mu + 332)(\mu - 3) C^2_2(G) C^2_2(R)} 
{(\mu - 1)} \right. \nonumber \\ 
&& \left. ~~~~~~~-~ 40(\mu - 3) C_2(G) C^2_2(R) T(R) \Nf ~+~ 
\frac{32(2\mu^2 - 15\mu + 19)(\mu - 3) d_A^{abcd} 
d_F^{abcd}}{(\mu - 1) N_{\mbox{\footnotesize{fund}}}} \right. \nonumber \\  
&& \left. ~~~~~~~-~ \frac{2(73\mu^2 - 462\mu + 545)(\mu - 3) C^3_2(G) C_2(R)} 
{3(\mu - 1)} \right. \nonumber \\
&& \left. ~~~~~~~+~ \frac{2(6\mu^2 - 24\mu + 19) C^2_2(G) C_2(R) T(R)\Nf} 
{(\mu - 1)} \right. \nonumber \\ 
&& \left. ~~~~~~~+~ \frac{96 d_F^{abcd} d_F^{abcd} \Nf} 
{(\mu - 1) N_{\mbox{\footnotesize{fund}}}} \right] 
\frac{(2\mu-1)\Gamma(4-3\mu)\Gamma^2(2-\mu)\Gamma(3\mu-2)\Gamma^5(\mu) g^3} 
{\Gamma(3-2\mu)\Gamma(4\mu-2)\Gamma^2(2\mu-1)} ~.  
\end{eqnarray}  
Whilst the diagram with a double operator insertion has the value  
\begin{eqnarray} 
&& i Z^2_{32} \left[ 20(2\mu^2 - 15\mu + 19)(\mu - 3) C_2(G) C^3_2(R) ~-~ 
8(2\mu^2 - 15\mu + 19)(\mu - 3) C^4_2(R) \right. \nonumber \\ 
&& \left. ~~~~~~~-~ \frac{33(2\mu^2 - 15\mu + 19)(\mu - 3) C^2_2(G) C^2_2(R)} 
{2} ~-~ \frac{ C^2_2(G) C_2(R) T(R) \Nf}{2} \right. \nonumber \\ 
&& \left. ~~~~~~~+~ \frac{55(2\mu^2 - 15\mu + 19)(\mu - 3) C^3_2(G) C_2(R)}{12}
{}~-~ \frac{8(2\mu^2 - 15\mu + 19)(\mu - 3) d_A^{abcd} d_F^{abcd}} 
{N_{\mbox{\footnotesize{fund}}}} \right. \nonumber \\
&& \left. ~~~~~~~-~ \frac{24 d_F^{abcd} d_F^{abcd} \Nf}
{N_{\mbox{\footnotesize{fund}}}} \right] 
\frac{\mu(2\mu-1)\Gamma(3-2\mu)\Gamma^3(\mu) g^2}{\Gamma(3\mu-1)} ~.  
\end{eqnarray} 
This diagram contributes when one considers the powers of the coupling constant
which are present. This completes the evaluation of all the diagrams relevant
for the renormalization of the two point function at four loops. 

\sect{Four loop anomalous dimension.} 

Having determined all the contributions to the two point function including 
those from evanescent operator insertions it is a relatively straightforward
task to determine the four loop correction to the wave function 
renormalization. We find 
\begin{eqnarray} 
Z_\psi^{\mbox{\footnotesize{$\MSbar$}}} &=& 1 ~-~ \frac{C_2(R) T(R)\Nf g^2} 
{4\pi^2\epsilon} ~-~ C_2(R) \left[ \frac{C_2(G) T(R)\Nf}{12\epsilon^2} \right. 
\nonumber \\ 
&& \left. ~~~~+~ [4T^2(R) \Nf^2 + C_2(G) T(R) \Nf + 2C^2_2(G) - 6C_2(R) C_2(G) 
+ 4C^2_2(R)] \frac{1}{24\epsilon} \right] \frac{g^3}{\pi^3} \nonumber \\ 
&& -~ \left[ \frac{T(R)\Nf C_2(R) C^2_2(G)}{32\epsilon^3} ~+~ 
\left( 192 \frac{d_A^{abcd} d_F^{abcd}}{N_{\mbox{\footnotesize{fund}}}} 
- 48 \frac{\Nf d_F^{abcd} d_F^{abcd}}{N_{\mbox{\footnotesize{fund}}}} 
- 96 C^3_2(G) C_2(R) \right. \right. \nonumber \\ 
&& \left. \left. ~~~~+ 352 C^2_2(G) C^2_2(R) + 3 C^2_2(G) C_2(R) T(R) \Nf 
- 448 C_2(G) C^3_2(R) \right. \right. \nonumber \\ 
&& \left. \left. ~~~~+ 32 C_2(G) C_2(R) T^2(R) \Nf^2 + 192 C^4_2(R) 
- 8 C^2_2(R) T^2(R) \Nf^2 \frac{}{} \right) \frac{1}{256\epsilon^2} \right. 
\nonumber \\
&& \left. ~~~~+ \left( 772 C^3_2(G) C_2(R) - 1536 \frac{d_A^{abcd} d_F^{abcd}} 
{N_{\mbox{\footnotesize{fund}}}} - 624 \frac{d_F^{abcd} d_F^{abcd} \Nf} 
{N_{\mbox{\footnotesize{fund}}}} \right. \right. \nonumber \\ 
&& \left. \left. ~~~~~~~~~~- 2808 C^2_2(G) C^2_2(R) + 299 C^2_2(G) C_2(R) T(R) 
\Nf + 3552 C_2(G) C^3_2(R) \right. \right. \nonumber \\ 
&& \left. \left. ~~~~~~~~~~- 864 C_2(G) C^2_2(R) T(R) \Nf - 1536 C^4_2(R) 
+ 576 C^3_2(R) T(R) \Nf \right. \right. \nonumber \\ 
&& \left. \left. ~~~~~~~~~~+ 192 C_2(R) T^3(R) \Nf^3 \frac{}{} \right) 
\frac{1}{1536\epsilon} \right] \frac{g^4}{\pi^4} ~+~ O(g^5) ~.  
\label{Zpsi} 
\end{eqnarray} 
Ordinarily one does not quote this value but merely gives the corresponding
anomalous dimension. We have done so here to illustrate an important feature of
renormalizing a model with evanescent operators. For ordinary (single coupling)
theories where this is not an issue the poles of the $\MSbar$ renormalization 
constants are related in a particular way. For example, the residues of the 
non-simple poles are determined by the simple pole residues at lower orders, 
\cite{47}. However, examining (\ref{Zpsi}) one can see that this is not the 
case since at four loops the new Casimir contributions occur in the residue of 
the double pole in $\epsilon$. Usually this would indicate an error in the 
renormalization since these new Casimirs cannot arise before four loops and so 
they should only appear in the simple pole in $\epsilon$. In this instance, 
however, the choice (\ref{Zpsi}) correctly renders the two point function 
finite but the anomalous dimension derived from it corresponds to the 
na\"{\i}ve renormalization group function and not the true one in relation to 
the discussion of section 2. To obtain the true anomalous dimension there are 
two possible procedures to follow. One is to extend the projection technique 
developed  in \cite{20,17} and applied to the $SU(\Nc)$ NATM $\beta$-function 
in \cite{15} to the present case. This would involve calculating the 
corrections to the projection functions by inserting each evanescent operator 
into a two point function and calculating the finite part after 
renormalization. These finite parts determine the projection function for each 
operator and together with the na\"{\i}ve $\beta$-functions of each operator, 
allows one to deduce the true renormalization group function. The alternative 
approach which produces the equivalent result is to ensure that the finite 
renormalized two point function in two dimensions satisfies the renormalization
group equation 
\begin{equation} 
\left[ {\tilde{\mu}} \frac{\partial ~}{\partial {\tilde{\mu}}} ~+~ \beta(g) 
\frac{\partial ~}{\partial g} ~+~ \frac{n}{2} \gamma(g) \right] 
\Gamma^{(n)}(p,{\tilde{\mu}},g) ~=~ 0 
\end{equation}  
where $\Gamma^{(n)}(p,\tilde{\mu},g)$ is the renormalized $n$-point Green's 
function. To ensure this condition can be used one must be careful in retaining
the finite parts of all the integrals in the construction of the Green's 
function before renormalization. This is the reason why we have been careful to
compute the integrals exactly in most cases and to sufficient powers in 
$\epsilon$ in the other cases and means we will follow the latter course here. 
After renormalization the two point function is 
\begin{eqnarray} 
\Gamma^{(2)}(p,\tilde{\mu},g) &=& \left[ 1 ~+~ C_2(R) \left( 4T(R)\Nf + C_2(G) 
- 2C_2(R) - 4T(R)\Nf \ln \left( \frac{p^2}{\hat{\mu}^2} \right) \right) 
\frac{g^2}{16\pi^2} \right. \nonumber \\ 
&& \left. ~+ \left( 12T(R)\Nf(C_2(G) - 4T(R)\Nf) \ln \left( 
\frac{p^2}{\hat{\mu}^2} \right) - 12 C_2(G) T(R)\Nf \ln^2 \! \left( 
\frac{p^2}{\hat{\mu}^2} \right) \right. \right. \nonumber \\
&& \left. \left. ~~~~~~~-~ 19C^2_2(G) ~+~ 66C_2(G)C_2(R) ~+~ 16C_2(G)T(R)\Nf ~-~
56 C^2_2(R) \right. \right. \nonumber \\ 
&& \left. \left. ~~~~~~~-~ 48 C_2(R) T(R)\Nf ~+~ 64 T^2(R) \Nf^2 \right) 
\frac{C_2(R)g^3}{192\pi^3} \right. \nonumber \\ 
&& \left. ~+ \left( 2112 \frac{d_A^{abcd} d_F^{abcd}} 
{N_{\mbox{\footnotesize{fund}}}} ~-~ 48C^2_2(G)C_2(R)T(R)\Nf \ln^3 \! \left( 
\frac{p^2}{\hat{\mu}^2} \right) \right. \right. \nonumber \\ 
&& \left. \left. ~~~~~~+~ 24 \left( 3C^2_2(G) - 20 C_2(G)T(R)\Nf 
+ 4 C_2(R)T(R)\Nf \right) \right. \right. \nonumber \\ 
&& \left. \left. ~~~~~~~~~~~~~~\times C_2(R) T(R)\Nf \ln^2 \! \left( 
\frac{p^2}{\hat{\mu}^2} \right) \right. \right. \nonumber \\ 
&& \left. \left. ~~~~~~+\, \left( 96C^3_2(R)T(R)\Nf ~-~ 44C^2_2(G)C_2(R)T(R)\Nf
{}~-~ 192C^2_2(R) T^2(R)\Nf^2 \right. \right. \right. \nonumber \\ 
&& \left. \left. \left. ~~~~~~~~~~~-~ 240C_2(G)C^2_2(R) T(R)\Nf ~+~ 
1152 C_2(G)C_2(R)T^2(R)\Nf^2 \right. \right. \right. \nonumber \\ 
&& \left. \left. \left. ~~~~~~~~~~~-~ 768C_2(R) T^3(R)\Nf^3 ~+~ 
2496 \frac{d_F^{abcd} d_F^{abcd}\Nf}{N_{\mbox{\footnotesize{fund}}}} \right) 
\ln \left( \frac{p^2}{\hat{\mu}^2} \right) \right. \right. \nonumber \\ 
&& \left. \left. ~~~~~~-~ 1147C^3_2(G)C_2(R) ~+~ 4452C^2_2(G) C^2_2(R) ~+~ 
1152 C_2(R) T^3(R)\Nf^3 \right. \right. \nonumber \\ 
&& \left. \left. ~~~~~~-~ 870 C^2_2(G)C_2(R) T(R) \Nf ~-~ 6096C_2(G) C^3_2(R)
\right. \right. \nonumber \\
&& \left. \left. ~~~~~~+~ 3408C_2(G) C^2_2(R) T(R) \Nf ~-~ 624C_2(G) 
C_2(R)T^2(R) \Nf^2 \right. \right. \nonumber \\
&& \left. \left. ~~~~~~+~ 2856C^4_2(R) ~-~ 2688C^3_2(R) T(R)\Nf ~-~ 
1056 C^2_2(R) T^2(R)\Nf^2 \right. \right. \nonumber \\
&& \left. \left. ~~~~~~+~ (576\zeta(3) - 3312) \frac{d_F^{abcd} d_F^{abcd}\Nf}  
{N_{\mbox{\footnotesize{fund}}}} \right) \frac{g^4}{3072\pi^4} \right] 
i \pslash ~+~ O(g^5) 
\end{eqnarray}  
where $\tilde{\mu}^2$ $=$ $4\pi e^{-\gamma}\hat{\mu}^2$ and $\gamma$ is the 
Euler-Mascheroni constant. Hence, using the three loop $\MSbar$ 
$\beta$-function of \cite{15} we find that for a general colour group the 
$\MSbar$ result is 
\begin{eqnarray} 
\gamma^{\mbox{\footnotesize{$\MSbar$}}}(g) &=& -~ C_2(R) T(R) \Nf 
\frac{g^2}{2\pi^2} \nonumber \\
&& +~ \left[ 2C_2(G) C_2(R) - C^2_2(G) - 2C_2(G) T(R)\Nf - 8T^2(R) \Nf^2 
\right] \frac{C_2(R)g^3}{16\pi^3} \nonumber \\ 
&& +~ \left[ 624 \frac{\Nf d_F^{abcd} d_F^{abcd}} 
{N_{\mbox{\footnotesize{fund}}}} + 57 C^3_2(G) C_2(R) - 198 C^2_2(G) C^2_2(R)
\right. \nonumber \\
&& \left. ~~~~~- 83 C^2_2(G) C_2(R) T(R) \Nf + 168 C_2(G) C^3_2(R) 
+ 144 C_2(G) C^2_2(R) T(R) \Nf \right. \nonumber \\ 
&& \left. ~~~~~- 192 C_2(R) T^3(R) \Nf^3 \right] 
\frac{g^4}{384\pi^4} ~+~ O(g^5) ~. 
\label{gammams} 
\end{eqnarray}  
As a check on this value we have in fact applied the projection formalism to
the case when $G$~$=$~$SU(\Nc)$ and verified that both are in agreement. One 
interesting feature of the expression (\ref{gammams}) is that the new Casimir 
$d_A^{abcd} d_F^{abcd}$ has cancelled in the final expression leaving only a 
term involving $d_F^{abcd} d_F^{abcd}$. This cancellation appears to be
consistent with our earlier discussion on the nature of the graphs in the two
point function when regarded from a QCD point of view and also reflects the
contribution from the relevant evanescent operators. Though we do not regard
this observation as a hard check on the final result. More importantly from the
point of view of our original motivation a new Casimir does appear in the 
($\MSbar$) anomalous dimension of the fermion.  

Whilst we have produced the $\MSbar$ anomalous dimension at four loops there
remains one final task to perform which has been discussed in related work. 
This rests in the nature of the dimensional regularization. In two dimensions
one has various equivalences between certain models such as the abelian 
Thirring model being the same as the Gross Neveu model for $\Nf$~$=$~$1$. This
is established through an elementary two dimensional Fierz transformation. In
$d$-dimensions, however, such relations cannot be preserved. In the first
instance, the Fierz transformation becomes infinite dimensional as one has to 
decompose $I \otimes I$ into the full basis of tensor products of 
$\Gamma_{(n)}$, \cite{33,34}. Secondly, the relation that one can establish
depends on the evanescent operators which are generated in the four point 
interaction but not in a way which the direct relationship can be determined.
Therefore the situation is such that in renormalizing the NATM, the resulting
$\MSbar$ renormalization group functions do not preserve the specific
equivalences. In other words taking the abelian limit of (\ref{gammams}) does
not recover the four loop result of \cite{22} for the $\Nf$ $=$ $1$ Gross Neveu
model. To ensure that this property is preserved in the renormalization having
been broken by the regularization, we need to make a finite scheme change. 
Such changes were considered in {\em all} previous work in this area, 
\cite{16,17,18,19}, and we proceed along similar lines to \cite{15} here.  

As the choice of possible scheme changes is infinite we restrict our study to
the class introduced in \cite{15}. There the scheme was changed from $\MSbar$ 
to one where the finite part was also absorbed into the renormalization 
constants. In order to examine the equivalences the finite part is parametrised
and constraints placed on the parameters by ensuring agreement of the 
renormalization group functions in the various limits. Whilst this may 
introduce a large degree of redundancy it allows for choices in future 
applications of the results. Here as we are dealing with a general group $G$ we
use the usual Casimirs as the basis for the parametrization of the finite part.
In particular we choose the finite part of $Z_\psi$ to be 
\begin{eqnarray} 
&& \CR \left[ w_{21} \CR + w_{22} \CG + w_{23} T(R) \Nf \right] 
\frac{g^2}{\pi^2} \nonumber \\ 
&& +~ \CR \left[ w_{31} C^2_2(R) + w_{32} C^2_2(G) + w_{33} T^2(R) \Nf^2 
+ w_{34} \CG \CR \right. \nonumber \\ 
&& \left. ~~~~~~~~~~~~ + w_{35} \CR T(R) \Nf + w_{36} \CG T(R) \Nf \frac{}{}
\right] \frac{g^3}{\pi^3} ~+~ O(g^4)  
\end{eqnarray}  
where the $\{w_{2n}\}$ and $\{w_{3n}\}$ are constants which can be easily 
related to the $SU(\Nc)$ choice of \cite{15}. It is worth noting that choosing 
these values one has to redo the renormalization systematically at each loop 
order as the $\epsilon$-pole structure of the new $Z_\psi$ becomes dependent on
the parameters. To determine the new renormalization group function in this 
general scheme one can either follow the general procedure for scheme changes 
given in, for example, \cite{28} which involves calculating the function which 
relates the coupling constants in both schemes or apply the method used to 
deduce (\ref{gammams}) from the renormalization group equation itself. In fact 
we used both methods to ensure that our manipulations are consistent. Hence, we
find that for the general scheme the anomalous dimension is 
\begin{eqnarray} 
\gamma^{\mbox{\footnotesize{gen}}}(\bar{g}) &=& -~ C_2(R) T(R) \Nf 
\frac{\bar{g}^2}{2\pi^2} \nonumber \\ 
&& -~ C_2(R) \left[ \left( b_{11} + \half \right) T^2(R) \Nf^2 
+ b_{13} C_2(R) T(R) \Nf + \left( w_{23} + b_{12} + \eighth \right) 
C_2(G) T(R)\Nf \right. \nonumber \\
&& \left. ~~~~~~~~~~~~+ \left( w_{21} - \eighth \right) C_2(G) C_2(R)  
+ \left( w_{22} + \sixteenth \right) C^2_2(G) 
\right] \frac{\bar{g}^3}{\pi^3} \nonumber \\ 
&& -~ \left[ \half \left( b^2_{11} + 3b_{11} + 2b_{21} + 1 \right) C_2(R) 
T^3(R) \Nf^3 \right. \nonumber \\ 
&& \left. ~~~~+ \left( b_{23} - 2 w_{23} + \threehalves b_{13} +  b_{11} b_{13}
\right) C^2_2(R) T^2(R) \Nf^2 \right. \nonumber \\ 
&& \left. ~~~~+ \left( b_{11} b_{12} + \threeeighths b_{11} 
+ \threehalves b_{12} +  b_{22} + w_{23} + \threehalves w_{33} \right) C_2(G) 
C_2(R) T^2(R) \Nf^2 \right. \nonumber \\ 
&& \left. ~~~~+ \left( b_{26} - 2w_{21} + \half b_{13}^2 \right) C^3_2(R) 
T(R) \Nf \right. \nonumber \\ 
&& \left. ~~~~- \left( 2 w_{22} - \threehalves w_{35} + \threeeighths - w_{21} 
- b_{25} - \threeeighths b_{13} - b_{12} b_{13} + \threeeighths b_{11} \right) 
C_2(G) C^2_2(R) T(R) \Nf \right. \nonumber \\ 
&& \left. ~~~~+ \left( \mbox{\small{$\frac{3}{2}$}} w_{36} 
+ \mbox{\small{$\frac{83}{576}$}} + w_{22} + b_{24} + \half b^2_{12} 
+ \threeeighths b_{12} 
+ \threesixteenths b_{11} \right) C^2_2(G) C_2(R) T(R) \Nf \right. \nonumber \\
&& \left. ~~~~+ \left( \mbox{\small{$\frac{3}{2}$}} w_{31} 
- \mbox{\small{$\frac{7}{16}$}} - \mbox{\small{$\frac{3}{8}$}} b_{13} 
\right) C_2(G) C^3_2(R) 
+ \left( \threehalves w_{34} + \mbox{\small{$\frac{33}{64}$}} 
+ \threesixteenths b_{13} - \threeeighths b_{12} \right) C^2_2(G) C^2_2(R) 
\right. \nonumber \\ 
&& \left. ~~~~+ \left( \threehalves w_{32} - \mbox{\small{$\frac{19}{128}$}} 
+ \threesixteenths b_{12} \right) C^3_2(G) C_2(R) 
- \frac{13d_F^{abcd} d_F^{abcd} \Nf}{8 N_{\mbox{\footnotesize{fund}}}} 
\right] \frac{\bar{g}^4}{\pi^4} ~+~ O(\bar{g}^5) 
\label{gammagen} 
\end{eqnarray} 
where $\bar{g}$ denotes the coupling constant of the new scheme and we have 
taken the finite part of the renormalization constant $Z_\psi^2 Z_g$ to be 
\begin{eqnarray} 
&& \left[ b_{11} T(R)\Nf + b_{12} \CG + b_{13} \CR \right] 
\frac{g}{\pi} \nonumber \\ 
&& +~ \left[ b_{21} T^2(R) \Nf^2 + b_{22} T(R) \Nf C_2(G) + b_{23} \CR T(R) \Nf 
\right. \nonumber \\ 
&& \left. ~~~~~+ b_{24} C^2_2(G) + b_{25} \CR \CG + b_{26} C^2_2(R) \right] 
\frac{g^2}{\pi^2} ~+~ O(g^3) ~.  
\end{eqnarray}  
Whilst we have discussed the constraints on the three loop parameters in 
\cite{15} we will redo that calculation here. This is partly to do with the 
fact that there were restrictions from two equivalences and for reasons already 
discussed we will only consider the one relating the abelian Thirring model and 
the Gross Neveu model. For this we recall that the abelian limit of the 
Casimirs is 
\begin{equation} 
C_2(R) ~\rightarrow~ 1 ~~,~~ C_2(G) ~\rightarrow~ 0 ~~,~~ T(R) ~\rightarrow~ 
1 ~~,~~ d_F^{abcd} d_F^{abcd} ~\rightarrow~ 1 ~.  
\label{abellim} 
\end{equation} 
Therefore, we find the constraints 
\begin{eqnarray} 
b_{11} ~+~ b_{13} &=& -~ \frac{1}{2} \nonumber \\ 
b_{21} ~+~ b_{23} ~+~ b_{26} ~-~ 2w_{21} ~-~ 2w_{23} &=& \frac{11}{8} 
\label{cons} 
\end{eqnarray} 
Whilst these do not restrict the values of the parameters substantially any
(numerical) choice must satisfy these equations for the two dimensional 
symmetry broken in $d$-dimensions to be established. The first equation of
(\ref{cons}) agrees with that of \cite{15} when the parameters are converted to
the case of $G$ $=$ $SU(\Nc)$.  

\sect{Discussion.} 

One of our initial motivations was to examine the appearance or otherwise of 
new Lie group Casimirs in the NATM fermion anomalous dimension which are known 
to arise at four loops in QCD. Whilst one of the two potential new Casimirs is 
absent in (\ref{gammams}) the one involving the fundamental representation 
remains. Therefore if the connection between the WZWN model and the NATM 
critical exponents is valid then the problem of how such a term cancels in the 
final critical exponent when approached from the renormalization group equation
point of view is still open. One possibility is that of choosing a 
renormalization scheme in such a way that this new term is absent from 
(\ref{gammams}). Although this would appear to resolve the issue it has 
implications for the other renormalization group functions as this scheme
choice impinges on how one renormalizes their associated Green's function. In
particular this will impact upon the $\beta$-function and it is possible they
will be transformed into it. Even if this were not the case, the computation of
$\beta(g)$ itself at four loops in the NATM will produce these new Casimirs as
well, which can be seen from examining the group theory of the Feynman diagrams
of the fermion four point function. This would appear to contradict the recent
exact $\beta$-function of \cite{41}. However, in that case it seems that the
use of conformal symmetry has somehow excluded the higher order Casimirs. A 
similar feature occurs in four dimensional gauge theories at four loops. If one
examines the four loop $\MSbar$ $\beta$-function of QCD, \cite{12}, for an 
arbitrary colour group the terms involving $d_F$ and $d_A$ will vanish in 
certain cases. For instance, when the quark is in the adjoint representation 
and $\Nf$ $=$ $\half$ then the resulting $\beta$-function contains no higher
order Casimirs. This is, of course, the result of the theory possessing a new
symmetry which is ${\cal N}$~$=$~$1$ supersymmetry. Thus, in the NATM it may be
that a similar mechanism such as the two dimensional conformal symmetry used in
the construction of \cite{41} is responsible and powerful enough to exclude the
$d_F^{abcd} d_F^{abcd}$ term in the NATM $\beta$-function to {\em all} orders
in a particular renormalization scheme. Though it is not clear in this approach
what form the all orders fermion anomalous dimension would take. Alternatively 
from the critical exponent point of view since $\beta(g)$ will contain 
information on the non-trivial fixed point, $g_c$, at which the renormalization
group functions are evaluated at to obtain the critical exponents, to examine 
the mechanics of the Casimir cancellation further would require the explicit 
form of the four loop NATM $\beta$-function in order to carry out a test of 
this point of view at this order of approximation. The complexity of such a 
calculation is on a footing equal to the renormalization of the two point 
function at {\em five} loops. Moreover, the four loop $\beta$-function of the 
usual simple Gross Neveu model, which would serve as a preliminary to a similar
calculation in the NATM, has yet to be performed. Therefore, to understand this
further would require a substantial amount of new calculations beyond those 
performed here. 

\vspace{1cm} 
\noindent 
{\bf Acknowledgements.} This work was carried out with the support of 
{\sc PPARC} through a Postgraduate Studentship (DBA) and an Advanced Fellowship 
(JAG). The calculations were performed with the help of the computer algebra
and symbolic manipulation programmes {\sc Form}, \cite{25}, and {\sc Reduce},
\cite{48}. 

\appendix 

\sect{Gram determinant.} 

In this appendix we briefly recall the application of the Gram determinant 
method of \cite{26} to relate a two loop self energy integral in 
$(d+2)$-dimensions to a set of $d$-dimensional integrals. The method relies on 
several properties. First, the $d$-dimensional measure of a (massless) Feynman 
integral can be related through 
\begin{equation} 
d^{d+2} x ~=~ \frac{2\pi x^2}{d} d^d x ~.  
\end{equation} 
Next the Gram determinant of three Lorentz vectors, $x$, $y$ and $z$, is 
defined by 
\begin{equation} 
\mbox{Gr}(x,y,z) ~=~ \det \left(  
\begin{array}{ccc} 
x^2 & xy & xz \\ 
yx & y^2 & yz \\
zx & zy & z^2 \\ 
\end{array} 
\right) 
\end{equation} 
and through considering the Lorentz invariance of each integration, one can
show that, \cite{22},  
\begin{eqnarray} 
&& \int d^d x \int d^d y \int d^d z ~ \mbox{Gr}(x,y,z) F(b_1 x^2 + b_2 y^2 + b_3
z^2) \nonumber \\  
&& =~ \frac{d(d-1)(d-2)}{(2\pi)^3} \int d^{d+2} x \int d^{d+2} y \int d^{d+2} 
z ~ F(b_1 x^2 + b_2 y^2 + b_3 z^2)
\label{grameqn} 
\end{eqnarray} 
where $F(x)$ is a general function. In our case it represents the Feynman 
parametized two loop self energy graph of figure 7 when converted into a vacuum
diagram. For this topology one can always write the integral as a linear 
combination of the three Lorentz invariants $x^2$, $y^2$ and $z^2$ where $y$
and $z$ represent the internal $k$ and $l$ momentum integrations and $x$ 
corresponds to an integration over the endpoint to produce the three loop 
vacuum bubble. Since the Gram determinant is invariant under the change of
variables used to produce the form of the argument of $F(x)$ in (\ref{grameqn}) 
one can undo this transformation and replace $F(x)$ by the two loop topology of
figure 7. As the final $x$-integration is then the same for both sides one can
relate the {\em values} of the integrals on both sides quite straightforwardly.
Indeed, expanding out the Gram determinant and expressing all the scalar 
products in terms of factors which appear in the denominator, one obtains the 
general result which is implicit in \cite{22}. We have 
\begin{eqnarray} 
\left. \frac{}{} \langle \alpha_1, \alpha_2, \alpha_3, \alpha_4, \alpha_5 
\rangle \right|_{\mu+1} &=& \left[ \frac{}{} [-1,0,0,0,-1] ~-~ 
[-1,-1,0,0,0] ~+~ [-1,0,-1,0,0] \right. \nonumber \\ 
&& \left. ~+~ [0,0,0,-1,-1] ~+~ [0,-1,0,-1,0] ~-~ [0,0,-1,-1,0] \right. 
\nonumber \\  
&& \left. ~+~ [0,-1,0,0,-1] ~+~ [0,0,-1,0,-1] ~-~ [0,0,0,0,-2] \right. 
\nonumber \\  
&& \left. ~-~ [0,0,0,0,-1]_{-} ~-~ [-1,0,0,-1,-1]_{+} ~+~ [-1,-1,0,-1,0]_{+}
\right. \nonumber \\ 
&& \left. ~+~ [-1,0,-1,-1,0]_{+} \,+\, [-1,0,-1,0,-1]_{+} \,+\,  
[-1,-1,-1,0,0]_{+} \right. \nonumber \\ 
&& \left. ~-~ [-1,0,-2,0,0]_{+} ~-~ [-2,0,-1,0,0]_{+} ~+~ [0,-1,0,-1,-1]_{+}
\right. \nonumber \\ 
&& \left. ~+~ [0,-1,-1,-1,0]_{+} ~-~ [0,-2,0,-1,0]_{+} ~-~ [0,-1,0,-2,0]_{+} 
\right. \nonumber \\
&& \left. ~-~ [0,-1,-1,0,-1]_{+} \frac{}{} \right] \frac{1}{2(\mu-1)(2\mu-1)} 
\label{gram} 
\end{eqnarray} 
where 
\begin{equation} 
[n_1,n_2,n_3,n_5,n_5]_{n_6} ~=~ 
\langle \alpha_1 + n_1, \alpha_2 + n_2, \alpha_3 + n_3, \alpha_4 + n_4, 
\alpha_5 + n_5 \rangle \frac{1}{(x^2)^{n_6}} 
\end{equation} 
and the left hand side of (\ref{gram}) represents the value of the graph of 
figure 7 but evaluated in $(d+2)$-dimensions.

\end{document}